\documentclass[preprint,12pt,3p]{elsarticle}

\biboptions{sort&compress}

\usepackage{amssymb,amsmath,amsthm}

\usepackage{lipsum}
\usepackage{amsfonts}
\usepackage{graphicx}
\usepackage{epstopdf}
\usepackage{algpseudocode}
\usepackage{tikz}
\definecolor{softred}{RGB}{225,40,40}
\newtheorem{theorem}{Theorem}%
\newtheorem{corollary}{Corollary}

\usepackage{algorithm}

\usepackage{hyperref}
\allowdisplaybreaks[4]
\definecolor{ccr}{RGB}{0,128,0}  
\makeatletter

\newcommand{\Rmnum}[1]{\expandafter\@slowromancap\romannumeral #1@}
\makeatother
\hypersetup
{
pdfauthor={},%
pdfstartview=FitH,
CJKbookmarks=true,%
bookmarksnumbered=true,
bookmarksopen=true,%
colorlinks=true,
linkcolor=ccr ,%
urlcolor=ccr,%
citecolor=ccr,%
}
\hypersetup{pdfauthor={Name}}

\journal{Information Fusion}

\begin{document}

\begin{frontmatter}



\title{Evolutionary dynamics of collective decision-making with local social influence on static and dynamic networks} 


\author[tong]{Yuyuan Liu}
\author[tong]{Xiaojie Chen\corref{cor1}}
\ead{xiaojiechen@uestc.edu.cn}
\cortext[cor1]{Corresponding author}
\affiliation[tong]
{
  organization={School of Mathematical Sciences},
  addressline={University of Electronic Science and Technology of China},
  state={Chengdu},
  postcode={611731}, 
  country={China}
}

\begin{abstract}
Collective decision-making is ubiquitous across the living world and artificial societies. Individuals often choose an option based on intrinsic values of options. However, individual decision-making is also swayed by neighbors' choices, generating local social influence. Hence, an important question arises naturally, yet remains unanswered: when such social influence is integrated into the individual evaluation process for option choices, how does it affect collective decision-making outcomes in structured populations modeled by graphs. To address this, we consider a baseline model of binary options with social influence and assume that individuals not only evaluate the intrinsic values of options, but are also influenced by their neighbors' choices. We propose a perceived utility function integrating these two aspects for individual decision-making. By means of theoretical analysis, we first derive the average frequency of an option on static weighted connected graphs and present the mathematical condition under which this option prevails in the population. We find that the introduction of social influence can amplify the advantage of a superior option or compensate for the deficiency of an inferior one. We also reveal that the average degree of network exerts a dual effect on collective decision outcomes. Furthermore, we consider our evolutionary model on dynamic networks switching among distinct graph configurations. Our theoretical analysis shows that the evolutionary outcomes depend not only on the average degree of each network configuration, but also on its expected duration. We perform computer simulations to verify our theoretical predictions on static and dynamic networks.
\end{abstract}


\end{frontmatter}

\section{Introduction}\label{Xsec1-1}
Collective decision-making is widespread and critically important both in human society and nature~\cite{bib0001,bib0002}. In human society, it is essential for a wide range of activities, ranging from democratic elections and public policy formulation to addressing climate change issues~\cite{bib0003,bib0004,bib0005}. In nature, collective decision-making is crucial to the survival of species. A classic example is nest-site selection by honeybees~\cite{bib0006}. Despite its ubiquity, the underlying mechanisms governing the phenomenon of collective decision-making are not yet fully elucidated, which remain a prominent focus attracting considerable attention in recent years~\cite{bib0007,bib0008,bib0009,bib0010,bib0011,bib0012}.

Recently, models of local rules that define the interactions of individuals with one another and with the external environment have attracted significant attention. These models have proved enormously useful in uncovering principled explanations for collective decision-making and in designing efficient protocols for group decisions in decentralized cyber-physical systems, such as swarm robotics~\cite{bib0013}. In particular, individual decision-making can be modeled as a dynamical process. In such a framework, individuals update their options via local interactions~\cite{bib0014,bib0015}. They leverage the observable choices to infer or learn the options' underlying values. Through processing available information, they usually make a decision to maximize their utility~\cite{bib0016}.  Building upon this foundation, different types of update procedures have been proposed to mimic how individuals change their options, inspired by animal or human behaviors of decision~\cite{bib0017,bib0018,bib0019}. The impact of these learning protocols on outcomes
of collective decision-making has been extensively studied to explore the underlying mechanisms~\cite{bib0017,bib0018}. For instance, Yang {\rm et al.} studied how a population of individuals makes a collective decision when some members learn from others instead of evaluating the options on their own, and found that if the proportion of social learners exceeds a critical threshold, an option may prevail regardless of its merit value~\cite{bib0017}. Subsequently, Li {\rm et al.} expanded this framework to analyze the interplay between individual and social learning on collective decision-making outcomes in arbitrary connected network topologies~\cite{bib0019}. They demonstrated that although the introduction of social learning impairs the collective performance of the population choosing the option with higher merit value, the majority of the population always settles on this superior option regardless of the preference of social learning for individuals.

It is worth mentioning that, however, individuals choose the option based solely on intrinsic values of options within the framework mentioned above~\cite{bib0017,bib0019}. In practice, individual decision-making is also swayed by the choices of surrounding individuals, generating local social influence~\cite{bib0020,bib0021,bib0022,bib0023,bib0024}. For example, previous research has shown that when faced with a local consensus that conflicts with personal information, individuals tend to conform to others' choices~\cite{bib0021}. Consequently, when making choices, individuals not only evaluate options' intrinsic values but are also simultaneously influenced by the choices of their neighbors. Yet, a central question remains unanswered: how does such social influence modulate the  collective decision-making outcome in a structured population of individuals?

To address this question, in this work we construct a binary-choice dynamical model with social influence. We consider a structured population where each individual chooses between two options, $A$ and $B$. When making choices, individuals evaluate options' intrinsic values,  while simultaneously being influenced by the choices of their neighbors. Correspondingly, we mathematically depict the social influence of neighbors choosing different options on the focal individual who needs to update the option choice. Furthermore, we introduce a perceived utility function integrating these two aspects. Individuals then update their choices through introspection dynamics~\cite{bib0025}. Under these settings, we quantify collective decision-making outcomes by using the average frequency of individuals choosing option $A$, for the sake of convenience. The primary contributions of this study are outlined as follows.

(1) We construct a theoretical framework to investigate the effect of social influence on collective selection in all connected network structures. Unlike prior works~\cite{bib0017,bib0019}, we assume that when making choices, individuals not only consider the intrinsic value of options, but are also influenced by their neighbors' choices. Furthermore, we mathematically analyze how social influence shapes collective outcomes by using the approach of discrete-time Markov process.

(2) We theoretically derive the average frequency of option $A$ in any static weighted connected network and present the conditions under which option $A$ becomes dominant. We find that, when neighbors with option $A$ exert a more substantial social influence strength, individuals tend to opt for it, despite its comparatively low intrinsic value. We also find that the network's weighted average degree has a dual effect on collective decision outcomes. Specifically, when an option exhibits strong social appeal, its frequency increases with the average degree. Conversely, when an option's social appeal is weak, an increase in average degree diminishes its frequency. Furthermore, we conduct simulations on four representative network structures, including regular, small-world, scale-free, and random networks, and find that the results verify our theoretical predictions.

(3) Besides static networks, we consider our model in dynamic networks. We accordingly obtain the condition under which option $A$ can prevail. In addition, we find that when the population structure switches between different network topologies over time, the collective decision outcomes depend not only on the network structures, but also on the expected duration of each network configuration. In particular, the option with an inherent disadvantage but strong social appeal can achieve a frequency advantage when the population spends long durations in networks with high average degrees.

The remainder of the paper is organized as follows. In Section \ref{section2}, we introduce the evolutionary model of binary options with social influence on static networks. In Section \ref{section3}, we theoretically analyze the impact of social influence on collective decision-making in static networks and present numerical simulations. In Section \ref{section4}, we consider our model on dynamic networks, theoretically analyze the impact of social influence on collective decision-making, and provide numerical simulations. Our conclusion and discussion are presented in Section \ref{section5}.




\section{Evolutionary model involving social influence and introspection dynamics}\label{section2}

We consider a population of $N$ individuals, where the population structure can be described by a weighted, undirected graph $G = (V, E)$, where each node represents an individual and the relationship between individuals is linked by edges. Let $V = \{1, 2, \dots, N\}$ denote the set of vertices corresponding to the $N$ individuals, and $E \subseteq V \times V$ represent the set of edges capturing their interactions. The structure of $G$ is defined by a weighted adjacency matrix $\mathbf{W} = (\omega_{ij})_{N \times N} \in \mathbb{R}^{N \times N}$~\cite{bib0026,bib0027}. $\omega_{ij} > 0$ denotes the weight of the edge between individuals $i$ and $j$ if they are connected, and $\omega_{ij} = 0$ otherwise (assuming no self-loops, i.e., $\omega_{ii}=0$). The weighted degree of an individual $ i $ is given by $d_i=\sum_{j} \omega_{ij}$. Since the graph is undirected, we have $\omega_{ij} = \omega_{ji}$ for any $i, j \in V$.

At each time step, each individual holds one of two options $A$ and $B$, which have intrinsic values $\Pi_{A}$ and $\Pi_{B}$. So the option state of the population can be represented by a vector $ \mathbf{s}= \left(s_{1}, s_{2}, \ldots, s_{N}\right) \in\mathcal{S}=\{0,1\}^{N}$ (the Cartesian product of $N$ sets $\{0,1\}$). In particular, we set $s_i=1$ if individual $i$ selects option $A$, and $s_i=0$ for option $B$. Furthermore, we denote the state of all individuals except individual $ i $ as $ \mathbf{s}_{-i}=\left(s_{1}, \ldots, s_{i-1}, s_{i+1}, \ldots, s_{N}\right) \in\hat{\mathcal{S}}=\{0,1\}^{N-1}$.

In the decision-making process, individuals evaluate options based on available information. In our model, we consider this evaluation to be driven by two distinct factors: the option's intrinsic value and the social influence exerted by neighbors who have already adopted it. To capture this evaluation, we introduce a perceived utility function. Specifically, for individual $i$, the perceived utilities of the options are formulated as
\begin{equation}\label{U}
\begin{aligned}
&U_{i}(A,\mathbf{s}_{-i})=\Pi_{A} +n_{A}(\mathbf{s}_{-i})k_{A}\Pi_{A},\\
&U_{i}(B,\mathbf{s}_{-i})=\Pi_{B} +n_{B}(\mathbf{s}_{-i})k_{B}\Pi_{B},
\end{aligned}
\end{equation}
where $n_{A}(\mathbf{s}_{-i})$ and $n_{B}(\mathbf{s}_{-i})$ respectively denote the sum of the weights of individual $i$'s neighbors who have chosen options $A$ and $B$. Specifically, $n_{A}(\mathbf{s}_{-i}) = \sum_{j\neq i} s_j\omega_{ij}$ and $n_{B}(\mathbf{s}_{-i}) = \sum_{j\neq i} (1-s_j)\omega_{ij}$. The parameters $k_{A}$ and $k_{B}$ $(k_{A}, k_{B} \geq 0)$ quantify the strength of social influence for  options $A$ and $B$, respectively~\cite{bib0028}. If $k_{A} = k_{B}=0$, social influence is absent, and decisions are based solely on the options' intrinsic values. In addition, we define the terms $k_{A}\Pi_{A}$ and $k_{B}\Pi_{B}$ as the "social appeal" of options $A$ and $B$, respectively. Consequently, the total effect of social influence experienced by an individual is the product of the option's social appeal and the total weight of neighbors who have selected that option. We emphasize that the rationale of such a multiplicative parameterization for social influence can reflect the real-world phenomena: when an option has a lower intrinsic value, the social influence associated with this option should be weak; conversely, when an option has a higher intrinsic value, the neighbors' choices can generate a stronger level of social influence.

\begin{figure*}[t!]
\centering
\includegraphics[width=0.95\textwidth]{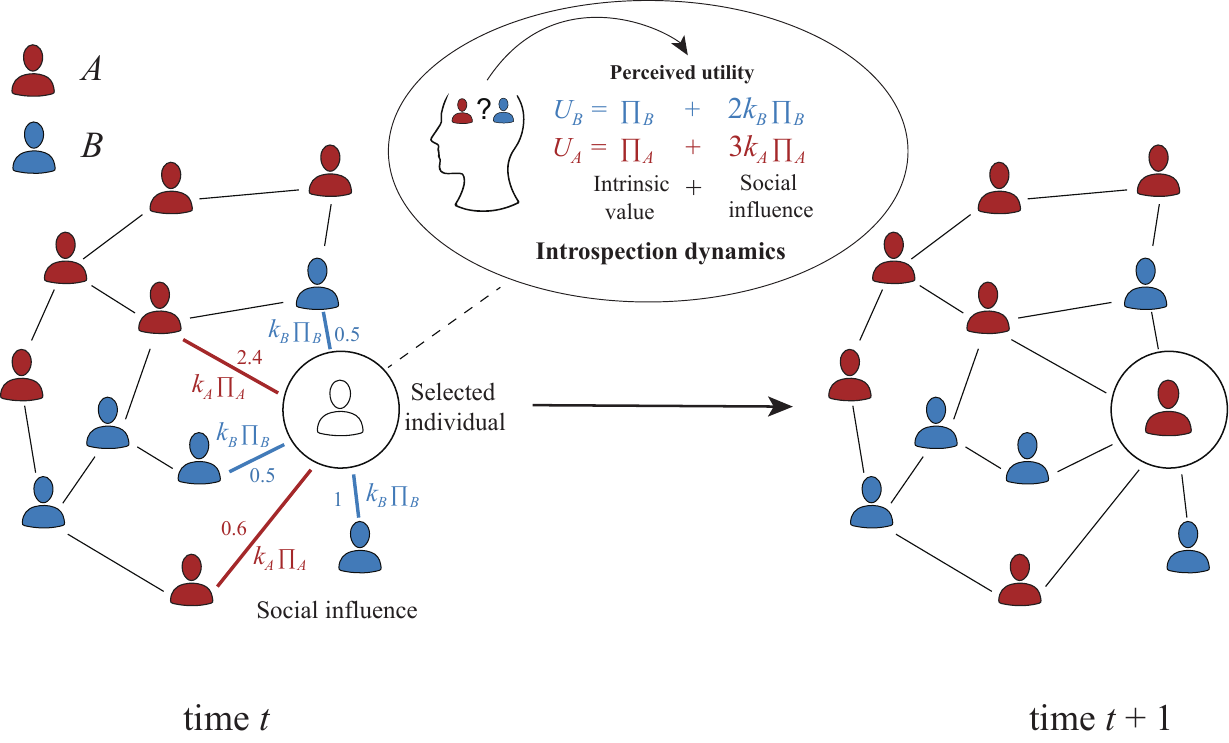}
\caption{
Schematic of  individual decision-making process under introspective dynamics. At time step $t$, a focal individual is randomly selected to re-evaluate its choice. This individual observes the current choices of its neighbors: red indicates neighbors who selected option $A$, while blue indicates those who selected option $B$. The collective choice of these neighbors constitutes the "social influence" on the focal individual, with $k_A$ and $k_B$ representing the strengths of social influence for each respective option. The focal individual evaluates the perceived utilities, $U_A$ and $U_B$, for each option. This utility comprises two components: the intrinsic values of options ($\Pi_{A}$ or $\Pi_{B}$) and the total effect of social influence. In the specific example illustrated, the focal individual has two neighbors choosing option $A$ (with edge weights $2.4$ and $0.6$). Thus, the perceived utility of option $A$ is $U_A = \Pi_{A} + 3k_A \Pi_{A}$. Similarly, the utility for option $B$ is $U_B = \Pi_{B} + 2k_B \Pi_{B}$. Subsequently, the focal individual updates its choice by comparing the perceived utilities $U_A$ and $U_B$, which determines the new option state at time $t+1$.
}
\label{t0}
\end{figure*}

Subsequently, individuals have the opportunity to update their options based on the perceived utility functions. In our model, individuals update their options over time according to introspection dynamics~\cite{bib0025}. At each time step, an individual $ i $ is chosen at random from the population. Suppose that individual $i$ currently adopts option $A$. The probability that it switches to option $B$ is given by the Fermi function
\begin{align*}
P_{A \rightarrow B}\left(\mathbf{s}_{-i}\right)=\frac{1}{1+e^{-\beta \left(U_{i}\left(B, \mathbf{s}_{-i}\right)-U_{i}\left(A, \mathbf{s}_{-i}\right)\right)}},
\end{align*}
otherwise, it keeps its current option~\cite{bib0029,bib0030,bib0031}. If individual $i$'s original choice is option $B$, it follows the same procedure as above to update the choice for the next time step. Here $\beta \in [0, \infty)$ denotes the selection intensity, which  characterizes an individual's sensitivity to the utility difference. When $\beta=0$, option updates are random, irrespective of the utility difference. For $\beta>0$, individuals are more likely to adopt the option with the higher perceived utility. To help readers intuitively understand the update process, we provide a schematic of the individual decision-making process in Fig.~\ref{t0}. In addition, the model parameters and their corresponding definitions are listed in Table~\ref{table}. In the following section, we will explore how the collective decision outcome is influenced by social influence on networks.

\begin{table}
\renewcommand{\arraystretch}{1.3}
\caption{Model parameters and definitions}
\centering%
\scalebox{1}{
\begin{tabular}{ll}
\hline
Parameter  &Meaning  \\
\hline
$s_i$&the option state of individual $i$\\
$\mathbf{s}$&the option state of the population\\
$\mathbf{s}_{-i}$&the option state of the population except individual $ i $\\
$\mathcal{S}$&the set of population option states\\
$\hat{\mathcal{S}}$&the set of population option states excluding one individual \\
$\Pi_{A}$&the intrinsic value of  option $A$\\
$\Pi_{B}$&the intrinsic value of  option $B$\\
$k_A$&the strength of social influence for  option $A$\\
$k_B$&the strength of social influence for  option $B$\\
$N$&the number of individuals in the population \\
$\beta$&the selection intensity \\
\hline
\end{tabular}}
\label{table}
\end{table}

\section{Collective decision-making outcomes in static networks}\label{Xsec3-3}\label{section3}
\subsection{Theoretical analysis}\label{Xsec4-3.1}
Since individual updates depend solely on the current population configuration, the evolution of the population option state can be described by a Markov chain~\cite{bib0025}. The state space $\mathcal{S} = \{0, 1\}^N$ consists of all $2^N$ possible option configurations of the population.  The transition of the option state is governed by a transition matrix $\mathbf{M}$ of dimension $2^N \times 2^N$. Let $ \mathbf{M}_{\mathbf{s}, \mathbf{s}^{\prime}} $ represent the transition probability from state $ \mathbf{s} $ to state $ \mathbf{s}^{\prime} $. We assume that only one individual updates the option at each time step. Consequently, transitions to a different state occur only between two adjacent states $\mathbf{s}$ and $\mathbf{s}^{\prime}$ with a Hamming distance of $\Vert \mathbf{s}-\mathbf{s}^\prime \Vert_{1}=1$. Let $l(\mathbf{s}, \mathbf{s}^{\prime})$ denote the component index where these two adjacent states differ. In particular, $l(\mathbf{s}, \mathbf{s}^{\prime})=j$ indicates that individual $j$ updates the option during the transition from state $\mathbf{s}$ to state $\mathbf{s}^{\prime}$. $ \mathbf{M}_{\mathbf{s}, \mathbf{s}^{\prime}} $ is then given by
\begin{align*}
\mathbf{M}_{\mathbf{s}, \mathbf{s}^\prime}
=\left\{
\begin{array}{ll}
 \hat{P}(\mathbf{s},\mathbf{s}^\prime)& \text { if } \Vert \mathbf{s}-\mathbf{s}^\prime \Vert_{1}=1, \\ 0 & \text { if } \Vert \mathbf{s}-\mathbf{s}^\prime \Vert_{1}>1,\\
1-\sum_{\hat{\mathbf{s}} \neq \mathbf{s}} \mathbf{M}_{\mathbf{s}, \hat{\mathbf{s}}} & \text { if } \mathbf{s}=\mathbf{s}^\prime.\end{array}\right.
\end{align*}
where $\hat{P}(\mathbf{s},\mathbf{s}^\prime)=\frac{1}{N} [s_{j}P_{A\rightarrow B}(\mathbf{s}_{-j})+(1-s_{j})P_{B\rightarrow A}(\mathbf{s}_{-j})]$ and $j=l(\mathbf{s}, \mathbf{s}^{\prime})$.

For any finite selection intensity $\beta\geq 0$, the transition probability between any two adjacent states is positive. Since any two states in $\mathcal{S}=\{0,1\}^N$ can be connected by a finite sequence of adjacent states, every state can be reached from any other state with positive probability within a finite number of time steps. Therefore, $\mathbf{M}$ is irreducible. Furthermore, under the above conditions, we have $ \mathbf{M}_{\mathbf{s}, \mathbf{s}} >0$ for any $\mathbf{s}\in\mathcal{S}$, and hence the Markov chain is aperiodic. Consequently, $\mathbf{M} $ has a unique stationary distribution. We denote this distribution by the vector $ \mathbf{x}=\left(x_{\mathbf{s}}\right)_{\mathbf{s} \in \mathcal{S}} $, which satisfies
\begin{align}\label{xM=x}
\mathbf{x}\mathbf{M}=\mathbf{x}, \quad \sum_{\mathbf{s} \in \mathcal{S}} x_{\mathbf{s}}=1.
\end{align}

Using Eq.~\eqref{xM=x}, we can derive an explicit expression for the stationary distribution as
\begin{align}\label{x=IXM}
\mathbf{x}=\mathbf{1}^{\boldsymbol{\top}}(\mathbf{I}+\mathbf{X}-\mathbf{M})^{-1},
\end{align}
where $\mathbf{1}$ denotes an all-ones vector with the same dimension as $\mathbf{x}$. $\mathbf{X}$ represents an all-ones square matrix with the same order as $\mathbf{M}$, and $\mathbf{I}$ is the identity matrix~\cite{bib0025}.

In particular, to further simplify Eq.~\eqref{x=IXM}, we introduce the function $\phi\left(\mathbf{s}\right): \mathcal{S} \to \mathbb{R}$. Let
\begin{align*}
\phi(\mathbf{s}) =~& N \Pi_{B}+\sum_{i}s_i\left(\Pi_{A}-\Pi_{B}\right) \\
&+\frac{1}{2}k_{A} \Pi_{A}  \sum_{i}\sum_{j} s_{i} s_{j} \omega_{ij} \\
&+\frac{1}{2}k_{B} \Pi_{B} \sum_{i}\sum_{j}\left(1-s_{i}\right)\left(1-s_{j}\right) \omega_{ij}.
\end{align*}
Then, the stationary distribution can be expressed as a function of $\phi$~\cite{bib0025}, given as
\begin{align}\label{x_expre}
x_{\mathbf{s}}=\frac{e^{\beta \phi(\mathbf{s})}}{\sum_{\mathbf{s}^{\prime} \in \mathcal{S}} e^{\beta \phi\left(\mathbf{s}^{\prime}\right)}}.
\end{align}
We prove that Eq.~\eqref{x_expre} is the unique solution of Eq.~\eqref{xM=x} (for more details, see Appendix A). In the subsequent section, we analyze the conditions favoring the dominance of option $A$ using the stationary distribution.

In the following, we calculate the expected frequency of option $A$ across the population. Let $f_{A}(\mathbf{s}) = \frac{1}{N} \sum_{i=1}^{N} s_{i}$ denote the fraction of individuals adopting option $A$ in a specific state $\mathbf{s}$. The average frequency of option $A$, $\bar{f_A}$, is obtained by using the expectation of $f_{A}(\mathbf{s})$ combined with the stationary distribution $\mathbf{x}$~\cite{bib0032}. Thus, $\bar{f_A}$ is given by
\begin{align*}
\bar{f_A}=\sum_{\mathbf{s} \in \mathcal{S}} f_{A}(\mathbf{s}) x_{\mathbf{s}}.
\end{align*}

To derive an explicit mathematical condition for the dominance of option $A$ (that is, $\bar{f_A}>1/2$), we calculate the expression of $\bar{f_A}$ in the weak selection limit ($\beta \rightarrow 0$). We emphasize that investigating the case of weak selection is justified since the perceived utility difference between two options is merely a part of the elements affecting the updating process in some scenarios~\cite{bib0019}. We then have the following conclusion.

\begin{theorem}\label{Theorem 1} When $\beta\rightarrow 0$, the average frequency of option $A$,  $\bar{f_A}$, in the static weighted connected network $G$, is given by
\begin{align}\label{main_ya}
\bar{f_A}&=\frac{1}{2}+\frac{\beta}{4}\left[\Pi_{A}-\Pi_{B}+\frac{\bar{d}}{2}\left(k_{A} \Pi_{A}-k_{B} \Pi_{B}\right)\right]+O\left(\beta^{2}\right),
\end{align}
where $ \bar{d}=\frac{1}{N} \sum_{i=1}^{N} d_{i} $ is the average weighted degree of the network $G$.
\end{theorem}

\noindent \textbf{Remark 1.} A detailed proof of Theorem \ref{Theorem 1} is provided in Appendix A. Theorem~\ref{Theorem 1} demonstrates that $\bar{f_{A}}$ depends on two key components: the intrinsic value difference $\Pi_{A}-\Pi_{B}$ and the product of the network's average weighted degree $\bar{d}$ and the social appeal difference $k_{A}\Pi_{A}-k_{B}\Pi_{B}$. Specifically, an increase in $\Pi_{A}$ enhances both the intrinsic advantage and the social appeal of option $A$, thereby monotonically increasing $\bar{f_{A}}$. Similarly, increasing $k_{A}$ promotes $ k_{A}\Pi_{A} $, favoring option $A$. Furthermore, increasing the network's average weighted degree $\bar{d}$ enhances the effect of social appeal. When $k_{A}\Pi_{A} > k_{B} \Pi_{B}$, larger $\bar{d}$ further promotes the adoption of option $A$.

Furthermore, according to Theorem \ref{Theorem 1}, we obtain the condition under which option $A$ is dominant, i.e., $ \bar{f_{A}} > 1/2$. It is described as follows.

\begin{corollary}
When $\beta\rightarrow 0$, $\bar{f_A} > 1/2$ if and only if
\begin{align}\label{main_ya_geq_0_2}k_{A} > k_{A}^{*} = \frac{2}{\bar{d}}\left(\frac{\Pi_{B}}{\Pi_{A}}-1\right) + k_{B}\frac{\Pi_{B}}{\Pi_{A}}.\end{align}
\end{corollary}

In the following, we perform numerical simulations to validate these theoretical predictions and further show the impact of social influence on collective outcomes.

\begin{algorithm}[t!]
    \caption{Simulation procedure for estimating $\bar{f}_A$}
    \label{alg:simulation}
    
    \footnotesize 
    \renewcommand{\baselinestretch}{0.8} 
    \selectfont 

\begin{algorithmic}[1]
        \Require population size $N$, intrinsic values $\Pi_A, \Pi_B$, social influence strengths $k_A, k_B$, selection intensity $\beta$, time steps $T$, number of network realizations $R$, and number of simulation runs $M$ for each network.
        \Ensure Estimated average frequency $\bar{f}_A$.

        \State $S \gets 0$

        \For{$r = 1$ to $R$}
            \State Generate a weighted adjacency matrix $\mathbf{W} = (\omega_{ij})_{N \times N}$ based on the given network model

            \For{$m = 1$ to $M$}
                \State $\mathbf{s} \gets$ an array of length $N$
                \For{$i = 1$ to $N$}
                    \State $s_i \gets$ a random integer from $\{0, 1\}$
                \EndFor

                \State $L \gets \emptyset$

                \For{$t = 1$ to $T$}
                    \State $i \gets$ a random integer from $\{1, \dots, N\}$
                    \State $n_A \gets 0$
                    \State $n_B \gets 0$

                    \For{$j = 1$ to $N$}
                        \If{$\omega_{ij} > 0$}
                            \If{$s_j = 1$}
                                \State $n_A \gets n_A + \omega_{ij}$
                            \Else
                                \State $n_B \gets n_B + \omega_{ij}$
                            \EndIf
                        \EndIf
                    \EndFor

                    \State $U_A \gets \Pi_A (1 + n_A k_A)$
                    \State $U_B \gets \Pi_B (1 + n_B k_B)$

                    \If{$s_i = 1$}
                        \State $\Delta U \gets U_B - U_A$
                    \Else
                        \State $\Delta U \gets U_A - U_B$
                    \EndIf

                    \State $P \gets 1 / (1 + \exp(-\beta \Delta U))$

                    \If{$\mathrm{rand}(0,1) < P$}
                        \State $s_i \gets 1 - s_i$
                    \EndIf

                    \State $c_A \gets 0$
                    \For{$q = 1$ to $N$}
                        \State $c_A \gets c_A + s_q$
                    \EndFor
                    \State append $c_A/N$ to $L$
                \EndFor

                \State $\bar{f}^{(r,m)}_A \gets \operatorname{Mean}(L)$
                \State $S \gets S + \bar{f}^{(r,m)}_A$
            \EndFor
        \EndFor

        \State $\bar{f}_A \gets S/(RM)$
        \State \Return $\bar{f}_A$
    \end{algorithmic}
\end{algorithm}

\begin{figure*}[t!]
\centering
\includegraphics[width=0.9\textwidth]{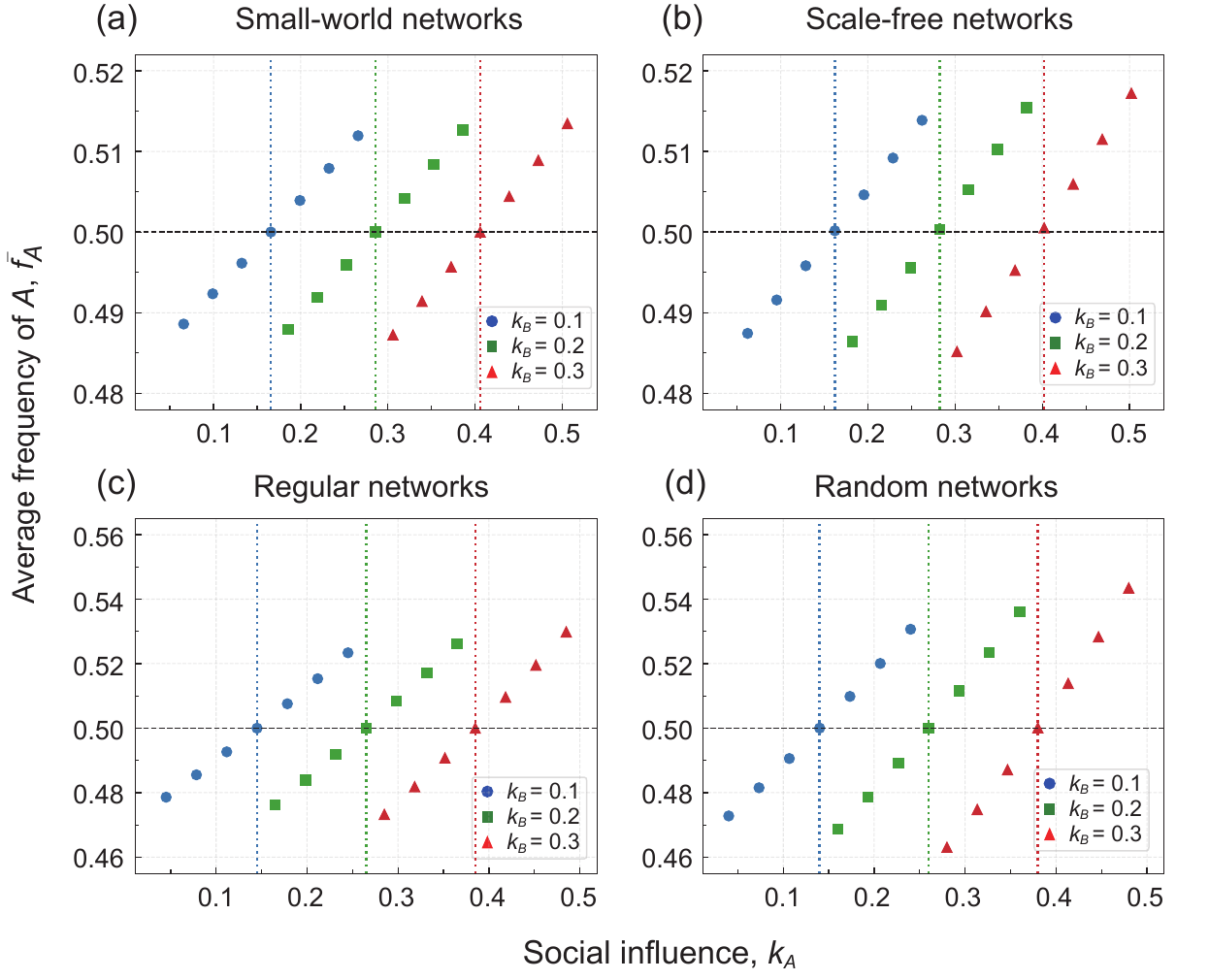}
\caption{The effect of social influence strength on collective decision-making outcomes. Panels (a)-(d) show the average frequency of option $A$, $\bar{f_A}$, as a function of its social influence strength, $k_{A}$, on small-world, scale-free, regular, and random networks, respectively. Each panel includes the results for three different values of social influence strength for option $B$: $k_{B}=0.1$ (blue circles), $k_{B}=0.2$ (green squares), and $k_{B}=0.3$ (red triangles). The horizontal dashed line indicates $\bar{f_A}=0.5$, while the vertical dashed lines denote the theoretical critical value $k_{A}^{*}$ at which $\bar{f_A}=0.5$ is met. Parameters are set as $N=100$, $\beta=0.01$, $\Pi_{A}=10$, and $\Pi_{B}=12$. Network configurations are: (a) NW small-world networks with initial degree $d=8$ and probability $p=0.1$ of connection addition; (b) BA scale-free networks with linking number $m=5$; (c) regular networks with degree $d=16$; and (d) random networks with $E=1000$ total edges. 
}
\label{t1}
\end{figure*}

\begin{figure*}[t!]
\centering
\includegraphics[width=0.9\textwidth]{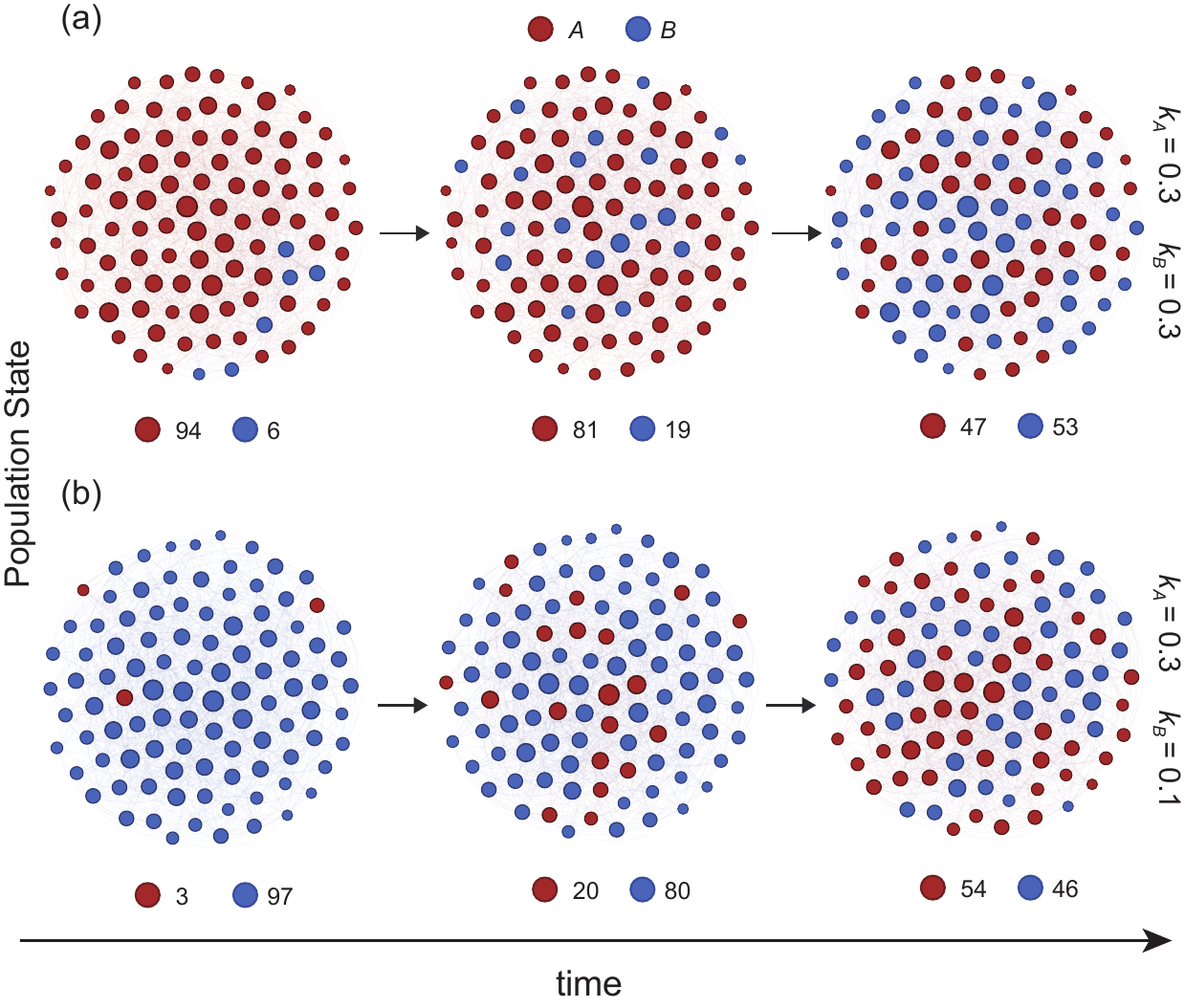}
\caption{Evolutionary spatial patterns of option states under social influence. Panels (a) and (b) present a series of snapshots from simulations under two different scenarios. In these snapshots, red nodes represent individuals choosing option $A$, while blue nodes denote individuals selecting option $B$. Node size is proportional to node degree (number of connections). Parameters are $N=100$, $\Pi_{A}=10$, $\Pi_{B}=12$, and $\beta=0.01$, where $k_{A}=k_{B}=0.3$ in panel (a) and $k_{A}=0.3$, $k_{B}=0.1$ in panel (b). The simulation program employs a random network generated with $E=1000$ total edges.}
\label{t2}
\end{figure*}

\begin{figure*}[t!]
\centering
\includegraphics[width=0.85\textwidth]{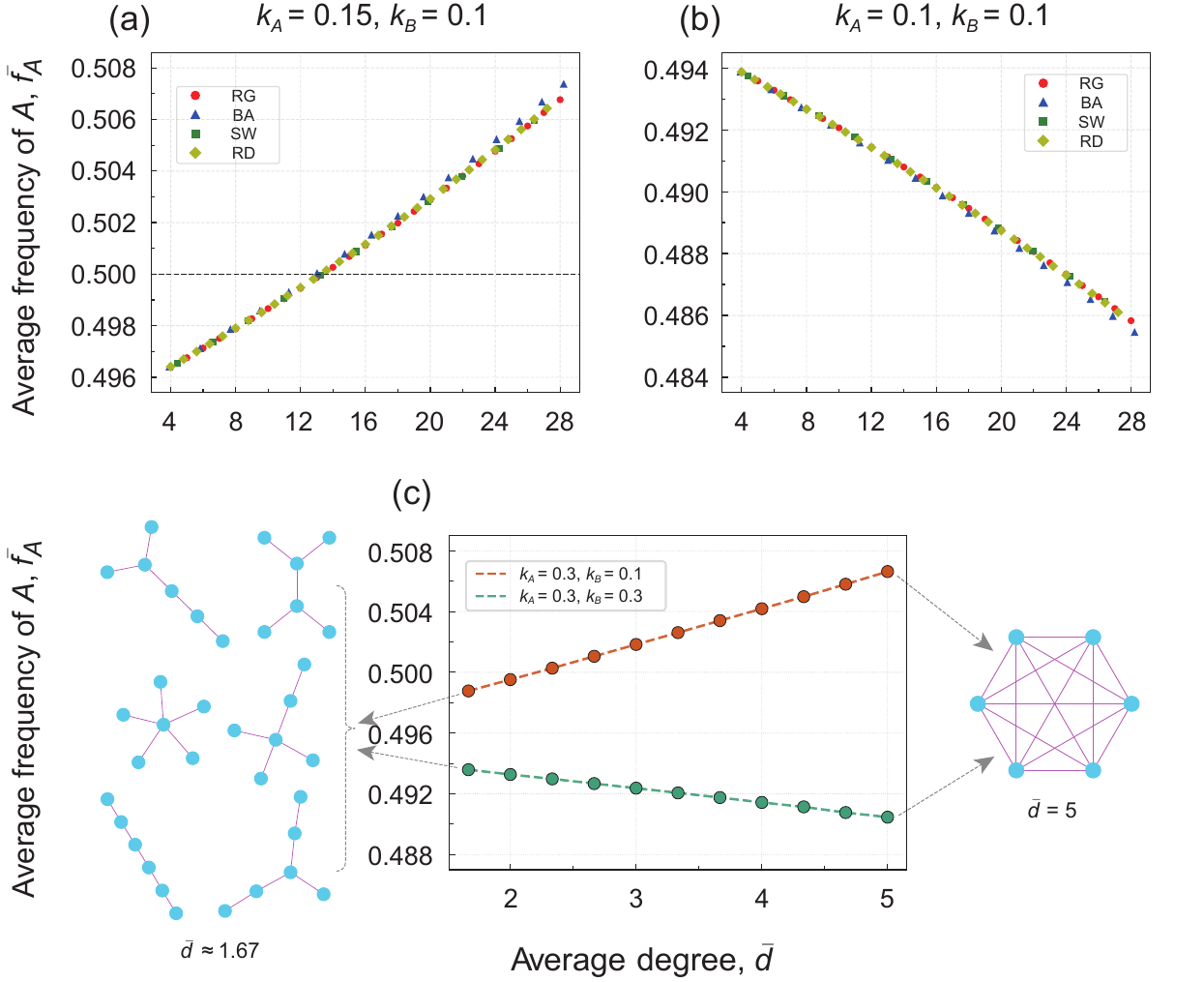}
\caption{The impact of network average degree on collective decision-making outcomes. Panels (a) and (b) illustrate the average frequency of option $A$, $\bar{f_A}$, in dependence on the network average degree $\bar{d}$ across the four types of networks: regular networks (RG, red circles), scale-free networks (BA, blue triangles), small-world networks (SW, green squares), and random networks (RD, yellow diamonds). Panel (c) shows the connected network topologies maximizing or minimizing $\bar{f_A}$ when the number of nodes is fixed. Parameters are set as $\beta=0.01$, $\Pi_{A}=10$, and $\Pi_{B}=12$. In panel (a), $k_{A}=0.15$, $k_{B}=0.1$, and $N=100$. In panel (b), $k_{A}=0.1$, $k_{B}=0.1$, and $N=100$. In panel (c), all unweighted connected graphs with $N=6$ are computed. The network topologies on which the average frequency of option $A$ is maximized and minimized are shown.
}
\label{t3}
\end{figure*}

\subsection{Numerical simulations}

In this subsection, we conduct Monte Carlo simulations on four representative networks, which are regular (RG), NW small-world (SW)~\cite{bib0033}, BA scale-free (BA)~\cite{bib0034}, and random (RD) networks. Simulations are performed on generated networks and begin from random initial distributions of options. The weight values of the network's edges are set to $0$ or $1$. For each type of network, we generate $30$ independent network realizations. For each generated network, we perform $10$ independent evolutionary simulation runs with different random initial option distributions. In each run, the system evolves for a total of $10^8$ time steps, and the  frequency of option $A$ is obtained by averaging over all these time steps, which is defined as  the evolutionary outcome in one simulation run. The final average frequency of option $A$ shown for static networks is then obtained by averaging over all $300$ simulation outcomes. More simulation details are provided in Algorithm~\ref{alg:simulation}.

We first show the role of social influence strength in the collective outcomes under weak selection ($\beta \rightarrow 0$). As shown in Fig.~\ref{t1}, we find a consistent trend across all four representative networks: while holding $k_{B}$ constant, the average frequency of option $A$, $\bar{f_A}$, increases monotonically with the social influence strength $k_{A}$. These results suggest that an increase in the social influence strength of one option can compensate for its disadvantage in intrinsic value, allowing for its dominance. Furthermore, a comparison of the curves in each panel reveals that when option $A$ is intrinsically inferior to $B$, a greater social influence strength is required for option $A$ to prevail ($k_{A} > k_{B}$, as predicted by Eq.~\eqref{main_ya_geq_0_2}). This numerical finding verifies our theoretical prediction.

To illustrate the impact of social influence on collective outcomes more intuitively, Fig.~\ref{t2} shows the evolutionary spatial patterns of population states in the case where option $B$ holds an intrinsic advantage over option $A$ ($\Pi_{A}=10<\Pi_{B}=12$). In Fig.~\ref{t2}(a),  although individuals adopting option $A$ initially have a quantity advantage, individuals gradually gravitate toward option $B$ over time. This trend indicates that when options $A$ and $B$ possess identical social influence strengths ($k_{A}=k_{B}=0.3$), the intrinsic value becomes the key factor influencing individual decision-making. Conversely, when option $A$ has a greater social influence than option $B$ ($k_{A}=0.3 > k_{B}=0.1$), individuals with option $A$ begin as the minority but eventually prevail in the population, as shown in Fig.~\ref{t2}(b). These results demonstrate that strong social influence strength can overcome the intrinsic disadvantage over time, favoring the dominance of option $A$.

\begin{figure*}[t!]
\centering
\includegraphics[width=0.9\textwidth]{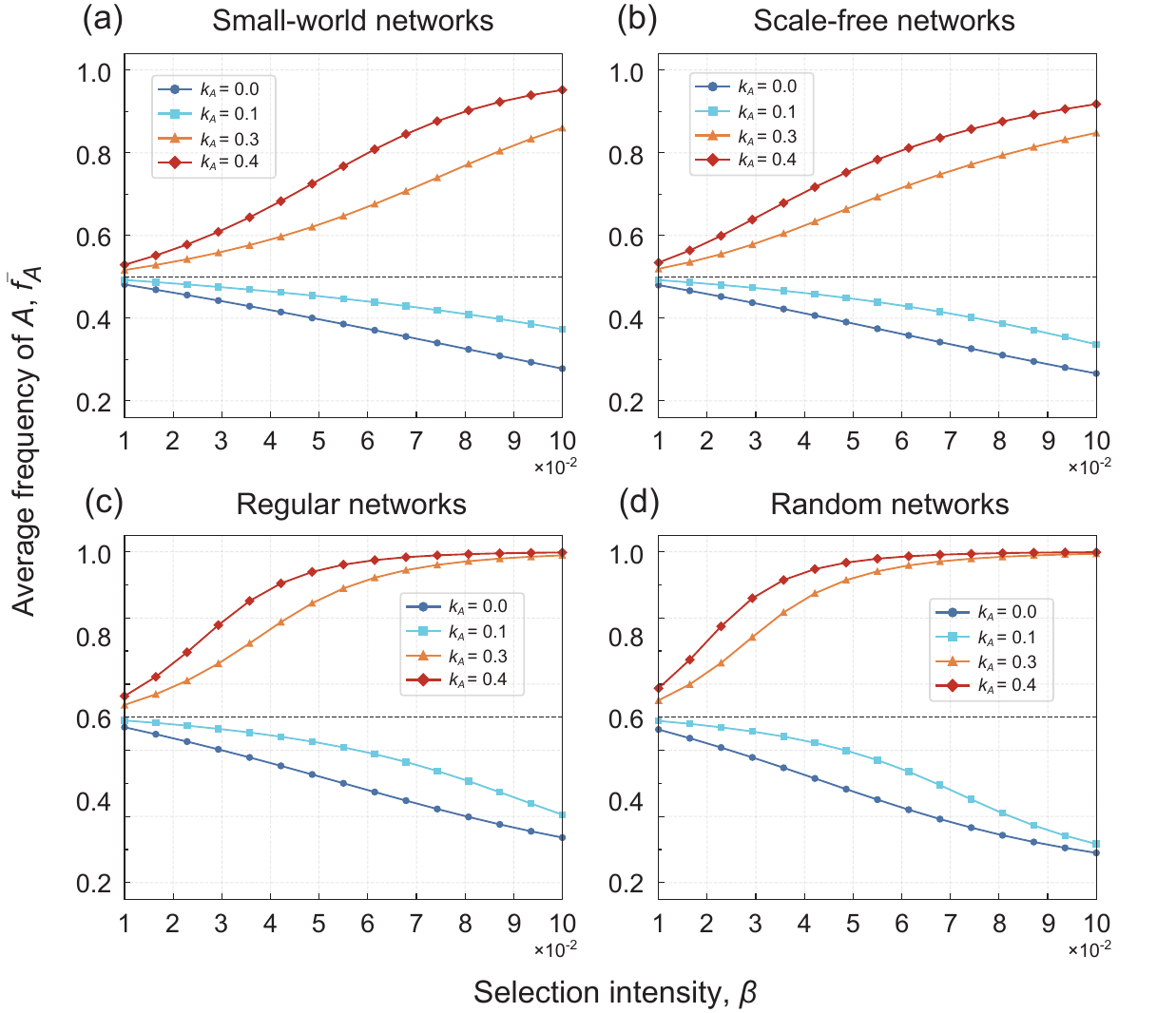}
\caption{The impact of selection intensity on collective decision-making outcomes under different social influence strengths. Panels (a)-(d) illustrate the average frequency of option $A$, $\bar{f_A}$, as a function of the selection intensity $\beta$ on the four types of networks. Each panel presents four different strengths of social influence for option $A$: $k_{A}=0.0$ (blue circles), $k_{A}=0.1$ (cyan squares), $k_{A}=0.3$ (orange triangles), and $k_{A}=0.4$ (red diamonds). Other parameters are set as $N=100$, $\Pi_{A}=10$, $\Pi_{B}=12$, and $k_{B}=0.1$. The network configurations are: (a) NW small-world networks with initial degree $d=8$ and probability $p=0.1$ of connection addition; (b) BA scale-free networks with linking number $m=5$; (c) regular networks with degree $d=16$; and (d) random networks with $E=1000$ total edges.
}
\label{t4}
\end{figure*}

We now analyze the influence of network average degree, $\bar{d}$, on collective outcomes. Fig.~\ref{t3} illustrates how the average frequency of option $A$ varies with the network's average degree, $\bar{d}$. As shown in Fig.~\ref{t3}(a), when $k_{A}\Pi_{A}>k_{B}\Pi_{B}$, the average frequency of option $A$ increases with the network's average degree $\bar{d}$ across all four types of networks. Conversely, Fig.~\ref{t3}(b) shows that when $k_{A} \Pi_{A} < k_{B} \Pi_{B}$, the average frequency of option $A$ monotonically decreases as $\bar{d}$ increases. These results indicate that the average degree acts as a key factor influencing collective outcomes.

In addition to the representative networks discussed above, we also examine which network structures, given a fixed number of nodes, favor or disfavor the prevalence of option $A$. We modulate the average degree by adjusting the total number of edges $E$ in the graphs of size $N=6$. For a given total number of edges $E$, we generate the set of all non-isomorphic connected graph structures. $\bar{f_A}$ is the mean outcome obtained by averaging $100$ independent realizations on each graph. Fig.~\ref{t3}(c) shows how the average frequency of option $A$ changes as the average degree of the networks increases from its lowest to its highest values. When option $A$ holds a social appeal advantage ($k_A\Pi_{A} > k_B\Pi_{B}$), the trend mirrors that of Fig.~\ref{t3}(a) (orange curve): the average frequency of $A$ increases monotonically with increasing the average degree. Specifically, the average frequency of $A$ is minimized in the sparsest networks, such as star and path graphs (see the left part of Fig.~\ref{t3}(c)), and is maximized in the fully connected complete graph (see the right part of Fig.~\ref{t3}(c)). Conversely, when option $A$ is disadvantaged ($k_A\Pi_{A} < k_B\Pi_{B}$), this relationship is inverted (green curve), which is consistent with the results shown in Fig.~\ref{t3}(b). In this regime, the sparsest networks sustain the highest frequency of $A$, whereas the complete graph suppresses it to its minimum. Taken together, these results demonstrate that whether increasing the network's average degree promotes or suppresses the prevalence of option $A$ is determined by the social appeal difference.

We present the above findings for the case of weak selection via theoretical analysis and numerical simulations. A natural question arising here is what happens when the selection intensity is not weak, since individuals tend to adopt the option with a higher utility in some scenarios.  However, as the selection intensity $\beta$ increases, analytical calculations become mathematically infeasible~\cite{bib0035}. In the following, we numerically examine the influence of $\beta$ on the collective outcomes. Fig.~\ref{t4} shows the average frequency of option $A$, $\bar{f_A}$, as a function of $\beta$ on the four different network structures. We find that when option $A$ is already favored ($\bar{f_A} > 0.5$) under small $\beta$ (orange and red curves), increasing $\beta$ further enhances its prevalence. Conversely, when option $A$ is disadvantaged under small $\beta$ ($\bar{f_A} < 0.5$, blue and cyan curves), a higher selection intensity accelerates the decline of $\bar{f_A}$. These results indicate that increasing the selection intensity can effectively magnify the prevalence or decline of option $A$, depending on whether option $A$ is dominated or not under weak selection.

\section{Collective decision-making outcomes in dynamic networks}\label{Xsec6-4}\label{section4}
The preceding investigation focuses on static population structures. In this section, we further explore the collective decision-making outcomes in dynamic networks as the population structure evolves over time~\cite{bib0036}.

\subsection{Individual decision process in dynamic networks}\label{Xsec7-4.1}

We first model the dynamic population structure as a time-varying graph. It switches among a finite set of $L$ distinct network configurations, denoted by $\mathcal{G}=\{G_{1}, G_{2}, \ldots, G_{L}\}$~\cite{bib0036}. For any index $\delta \in \mathcal{L} = \{1, 2, \ldots, L\}$, the network topology $G_{\delta}$ is defined by a specific weighted adjacency matrix $\mathbf{W}^{(\delta)}=(\omega_{ij}^{(\delta)})_{N \times N}$. The weighted degree of individual $i$ in network configuration $G_{\delta}$ is denoted by $d_{i}^{(\delta)}=\sum_{j} \omega_{ij}^{(\delta)}$, and the corresponding average weighted degree of the network is $\bar{d}^{(\delta)} = \frac{1}{N} \sum_{i=1}^N d_{i}^{(\delta)}$.

We assume that the individual decision update occurs prior to the network update~\cite{bib0036}. First, we analyze the individual decision-making process within a fixed network topology. We define the perceived utility function $U$ for individual $i$ in the current network $G_{\delta}$ as
\begin{align}
&U_{i}^{(\delta)}\left(A, \mathbf{s}_{-i}\right) = \Pi_{A} + n_{A}^{(\delta)}\left(\mathbf{s}_{-i}\right)  k_{A}\Pi_{A},\notag\\
&U_{i}^{(\delta)}\left(B, \mathbf{s}_{-i}\right) = \Pi_{B} + n_{B}^{(\delta)}\left(\mathbf{s}_{-i}\right)  k_{B}\Pi_{B},
\end{align}
where $n_{A}^{(\delta)}(\mathbf{s}_{-i}) = \sum_{j\neq i} s_j \omega_{ij}^{(\delta)}$ and $n_{B}^{(\delta)}(\mathbf{s}_{-i}) = \sum_{j\neq i} \left(1-s_j\right)\omega_{ij}^{(\delta)}$ denote the cumulative weight of neighbors adopting option $A$ and $B$ in network $G_{\delta}$, respectively. Subsequently, individuals update their options under the current network $G_{\delta}$. Individual $i$ with option $A$ switches to option $B$ with probability given by
\begin{align}
P_{A \rightarrow B}^{(\delta)}\left(\mathbf{s}_{-i}\right) = \frac{1}{1+e^{-\beta\left(U_{i}^{(\delta)}\left(B, \mathbf{s}_{-i}\right)-U_{i}^{(\delta)}\left(A, \mathbf{s}_{-i}\right)\right)}}.
\end{align}
Otherwise, it keeps its current option. If individual $i$'s choice is option $B$ currently, it follows the same procedure as above to update the choice by switching to $A$ or keeping $B$ for the next time step.

Following individual option updates, the network structure evolves according to a specific switching rule. In the subsequent section, we detail this network update process.

\subsection{Network transition process}\label{Xsec8-4.2}
We assume that the update of the network structure is an exogenous process, independent of the option state in the population. Specifically, the update of the network depends solely on its current configuration, allowing it to be modeled as an exogenous Markov chain~\cite{bib0036}. The state space $ \mathcal{L}=\{1,2, \ldots, L\} $ consists of the indices of the $ L $ $(L>1)$ possible network configurations, where each $ \delta \in \mathcal{L} $ corresponds to a network $ G_{\delta} $. The network transitions are governed by an $L \times L$ transition matrix $\mathbf{Q}=\left(q_{\delta \gamma}\right)$, where $q_{\delta \gamma}$ denotes the switching probability from network $G_{\delta}$ to $G_{\gamma}$.

Furthermore, we define $t_\delta$ as the expected duration of network $G_\delta$ in the population. Specifically, $t_\delta$ corresponds to the expected number of individual decision steps that occur when the population structure remains fixed at $G_\delta$~\cite{bib0036}. Upon leaving the current configuration $G_\delta$, we assume that the system switches to any of the remaining $L-1$ networks with equal probability. Given that the probability of exiting state $G_\delta$ in a single time step is $1/t_\delta$, the transition probabilities $q_{\delta \gamma}$ are defined as
\begin{align}\label{q}
q_{\delta \gamma}=\begin{cases}
1-\frac{1}{t_{\delta}} & \text { if } \delta=\gamma, \\
\frac{1}{t_{\delta}(L-1)} & \text { if } \delta \neq \gamma.
\end{cases}
\end{align}

Provided that $1\leq t_\delta<\infty$ for all $\delta\in\mathcal{L}$, and $t_{\delta_0}>1$ for at least one $\delta_0\in\mathcal{L}$, the Markov chain is irreducible and aperiodic, and hence ergodic. Thus, it possesses a unique stationary distribution, denoted by the vector $\mathbf{y}=(y_\delta)_{\delta \in \mathcal{L}}$. $\mathbf{y}$ is the unique solution to the following equation:
\begin{align}\label{yQ=y}
\mathbf{y}\mathbf{Q}=\mathbf{y}, \quad \sum_{\delta \in \mathcal{L}} y_{\delta}=1.
\end{align}
Solving Eq.~\eqref{yQ=y} with Eq.~\eqref{q}, we obtain a concise analytical form for the stationary distribution:
\begin{align}\label{yt}
y_\delta = \frac{t_\delta}{\sum_{k=1}^L t_k}.
\end{align}

\subsection{Theoretical analysis}\label{Xsec9-4.3}
In the following, we explore the average frequency of option $A$ in the dynamic network. To do that, we define the coevolution of collective decision and network topology on an extended state space $\Omega = \mathcal{S} \times \mathcal{L}$. This space comprises $L \cdot 2^N$ distinct states, each represented by an ordered pair $(\mathbf{s}, \delta)$, where $\mathbf{s} \in \mathcal{S}$ denotes the population's option configuration and $\delta \in \mathcal{L}$ represents the index of the current network structure. Since the transition from the current state $(\mathbf{s}, \delta)$ to a future state $(\mathbf{s}^{\prime}, \gamma)$ depends solely on the current configuration, the coevolutionary dynamics of collective decision and network structure can be depicted by a joint Markov chain.

Let $\mathbf{H}$ denote the transition matrix of this process, where the element $\mathbf{H}_{(\mathbf{s}, \delta), (\mathbf{s}^{\prime}, \gamma)}$ represents the transition probability from state $(\mathbf{s}, \delta)$ to $(\mathbf{s}^{\prime}, \gamma)$. Given that the network update process is assumed to be independent of the option state $\mathbf{s}$, this transition probability can be decomposed into the product of the option updating probability and the network switching probability. Thus, the transition probability $\mathbf{H}_{(\mathbf{s}, \delta), (\mathbf{s}^{\prime}, \gamma)}$ is given by
\begin{align*}
\mathbf{H}_{(\mathbf{s}, \delta), (\mathbf{s}^{\prime}, \gamma)} = \mathbf{M}_{\mathbf{s}, \mathbf{s}^{\prime}}^{(\delta)} \cdot q_{\delta \gamma},
\end{align*}
where $\mathbf{M}_{\mathbf{s}, \mathbf{s}^{\prime}}^{(\delta)}$ denotes the transition probability from state $\mathbf{s}$ to $\mathbf{s}^{\prime}$ under the specific network $G_{\delta}$.

The Markov chain defined by $\mathbf{H}$ is irreducible, as any state is reachable from any other within a finite number of steps. Furthermore, for any finite $\beta$, we have
$\mathbf{M}_{\mathbf{s},\mathbf{s}}^{(\delta)}>0$ for all
$\mathbf{s}\in\mathcal{S}$ and $\delta\in\mathcal{L}$. Moreover, suppose that there exists at least one $\delta_0\in\mathcal{L}$ such that $t_{\delta_0}>1$. Then,
$q_{\delta_0\delta_0}>0$. Hence, for any $\mathbf{s}\in\mathcal{S}$, $\mathbf{H}_{(\mathbf{s},\delta_0),(\mathbf{s},\delta_0)}
=\mathbf{M}_{\mathbf{s},\mathbf{s}}^{(\delta_0)}q_{\delta_0\delta_0}>0$. Thus, the Markov chain is
ergodic. Consequently, $\mathbf{H}$ has a unique stationary distribution, denoted by the vector $\mathbf{z}=\left(z_{(\mathbf{s}, \delta)}\right)_{(\mathbf{s}, \delta) \in \Omega}$. Here, $z_{(\mathbf{s}, \delta)}$ represents the long-term probability that the system stays in the option state $\mathbf{s}$ and the network structure is $G_{\delta}$. Besides, $\mathbf{z}$ is the unique solution to the following equation:
\begin{align}\label{zH=z}
\mathbf{z}\mathbf{H} = \mathbf{z}, \quad \sum_{(\mathbf{s}, \delta) \in \Omega} z_{(\mathbf{s}, \delta)} = 1.
\end{align}

To quantify the long-term collective outcome in dynamic networks, we calculate the expected frequency of option $A$. This quantity, denoted by $\langle f_A \rangle$, is obtained by averaging the instantaneous frequency $f_{A}(\mathbf{s})$ combined with the joint stationary distribution $\mathbf{z}$. This expectation is expressed as the sum of the frequency in each state weighted by its long-term probability, given by
\begin{align}\label{ffz}
\langle f_A \rangle=\sum_{(\mathbf{s}, \delta) \in \Omega} f_{A}(\mathbf{s}) z_{(\mathbf{s}, \delta)}.
\end{align}

To derive an analytical condition for the dominance of option $A$, we calculate the expression of $\langle f_A \rangle$ in the weak selection limit and obtain the following conclusion.

\begin{theorem}\label{Theorem 2} When $\beta\rightarrow 0$, the average frequency of option $A$,  $\langle f_A \rangle$, in dynamic networks is given by
\begin{align}
\langle f_A \rangle=~&\frac{1}{2}+\frac{\beta}{4}\left[\left(\Pi_{A}-\Pi_{B}\right)+\frac{\langle d\rangle}{2}\left(k_{A} \Pi_{A}-k_{B} \Pi_{B}\right)\right]+O\left(\beta^{2}\right),\label{dy_fa}
\end{align}
where
\begin{align}\label{d=yd}
\langle d\rangle=&\sum_{\delta=1}^{L} y_{\delta} \bar{d}^{(\delta)}=\sum_{\delta=1}^{L}\frac{t_\delta}{\sum_{k=1}^L t_k}\bar{d}^{(\delta)}.
\end{align}
\end{theorem}

\noindent \textbf{Remark 1.} A detailed derivation for Theorem~\ref{Theorem 2} is provided in Appendix A. Compared to Theorem~\ref{Theorem 1}, in dynamic networks one factor influencing the average frequency of $A$ is the weighted average degree $\langle d\rangle$, which depends on the stationary distribution of each network. Besides, the stationary distribution is determined by the expected duration of each network configuration. In particular, when the system is fixed on a single network $G_{\delta}$ (i.e., $y_\delta=1$), Theorem~\ref{Theorem 2} consistently reduces to Theorem~\ref{Theorem 1}.

Furthermore, according to Theorem~\ref{Theorem 2}, we obtain the condition under which option $A$ is dominant, i.e., $ \langle f_A \rangle > 1/2$. It is described as follows.

\begin{corollary}
When $\beta\rightarrow 0$, $\langle f_A \rangle > 1/2$ if and only if
\begin{align}k_{A} > \frac{2}{\langle d \rangle}\left(\frac{\Pi_{B}}{\Pi_{A}}-1\right) + k_{B}\frac{\Pi_{B}}{\Pi_{A}}.\end{align}
\label{Xenun4-2}
\end{corollary}

\subsection{Numerical simulations}\label{Xsec10-4.4}
In the following, we conduct numerical simulations to validate the theoretical predictions derived above and further explore the impact of social influence on collective outcomes in dynamic networks.

We first show the relationship between the weighted average degree and the expected durations of the networks. To do that, we consider two network configurations, $G_1$ with $200$ unweighted edges and $G_2$ with $1000$ unweighted edges, which have expected durations $t_1$ and $t_2$, respectively. We denote the average degree of networks $G_1$ and $G_2$ as $\bar{d}^{(1)}$ and $\bar{d}^{(2)}$, respectively, and have $\bar{d}^{(1)}=4$ and $\bar{d}^{(2)}=20$. In addition, according to Eq.~\eqref{d=yd}, we can obtain an expression for $\langle d\rangle$ in terms of $t_1$ and $t_2$, given by
\begin{align}\label{y1y2}
\langle d\rangle = \frac{t_1}{t_1 + t_2}\bar{d}^{(1)}+\frac{t_2}{t_1 + t_2}\bar{d}^{(2)}.
\end{align}

\begin{figure}[t!]
\centering {\includegraphics[width=11.8cm]{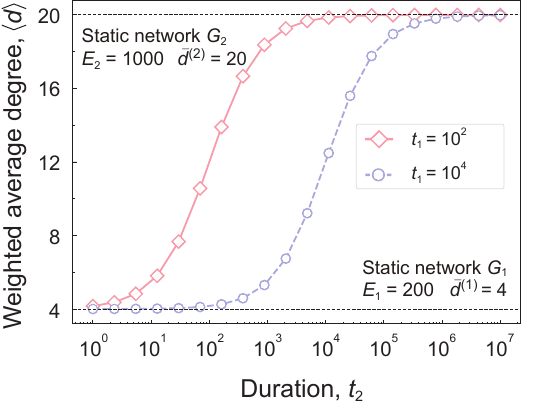}}
\caption{The dependence of the average degree $\langle d\rangle$ on the expected duration $t_2$ for two different values of $t_1$. The pink and purple curves correspond to $t_{1}=10^{2}$ and $t_{1}=10^{4}$, respectively. The two black horizontal dashed lines indicate the average degrees of the two static random networks, $G_{1}$ and $G_{2}$, respectively. The network configurations include the random network $G_{1}$ with $200$ edges and the random network $G_{2}$ with $1000$ edges.}
\label{t5}
\end{figure}

\begin{figure}[t!]
\centering {\includegraphics[width=12.8cm]{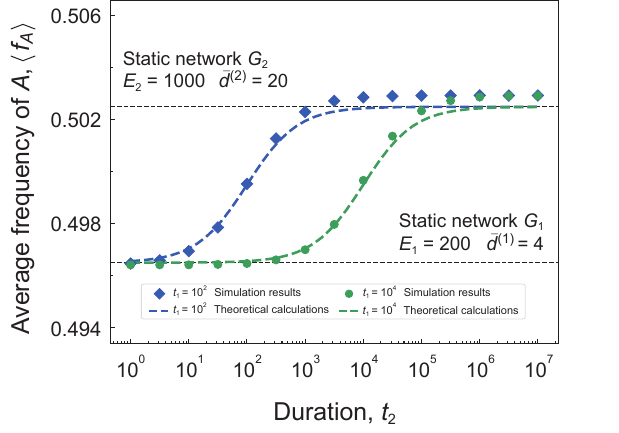}}
\caption{The average frequency of option $A$, $\langle f_A \rangle$, in dynamic networks. Blue diamonds (with a blue dashed line) correspond to the case $t_{1}=10^{2}$, and green circles (with a green dashed line) correspond to $t_{1}=10^{4}$. Markers show Monte Carlo simulation results and dashed lines show the corresponding theoretical predictions. For comparison, two black dashed lines are presented to indicate the average frequencies of $A$ on the network configuration $G_1$ and $G_2$, respectively. The parameters are $N = 100$, $\Pi_{A} = 10$, $\Pi_{B} = 12$, $k_A = 0.15$, $k_B = 0.1$, and $\beta = 0.01$. The simulation results are averaged over $100$ independently generated network pairs, where each pair consists of one random network with $200$ edges and another with $1000$ edges.
}
\label{t6}
\end{figure}

Fig.~\ref{t5} illustrates that when $t_{1}$ is fixed, increasing $t_{2}$ makes $\langle d\rangle$ gradually increase from $\bar{d}^{(1)}=4$ to $\bar{d}^{(2)}=20$. A smaller $t_{2}$ implies that the system spends more time in the sparser network $G_{1}$, leading to a larger $y_{1}$ and making $\langle d\rangle$ closer to $\bar{d}^{(1)}$. Conversely, a larger $t_{2}$ indicates a longer duration in the network $G_{2}$, resulting in a larger $y_{2}$, so that $\langle d\rangle$ is closer to $\bar{d}^{(2)}$. However, when $t_{1}$ is large, the system tends to remain in the sparse network for longer durations. As a result, achieving the same target value of $\langle d\rangle$ requires a proportionally larger $t_{2}$.

Based on the relationship between the expected duration parameters and weighted average degree, we further explore how the average frequency of option $A$ depends on the expected duration parameters in dynamic networks. We perform simulations over $100$ independent realizations. In each realization, we construct a specific pair of random networks, denoted as $G_1$ and $G_2$, configured identically to the setup in Fig.~\ref{t5}. The initial network configuration is set to $G_1$. The evolution proceeds in discrete time steps, where each step involves two stages: At each time step, an individual is selected to update its option on the current network topology $G_\delta$. Subsequently, the network topology switches to another network configuration or remains unchanged for the next time step. Specifically, the current network index $\delta$ updates to a new index $\gamma$ with probability $q_{\delta \gamma}$ as defined in Eq.~\eqref{q}. This two-stage process is repeated until the total simulation time of $10^9$ steps is reached. All other model parameters are the same as those used in Fig.~\ref{t3}. The final results are averaged over $100$ network realizations.

Accordingly, Fig.~\ref{t6} shows the average frequency of option $A$, $\langle f_A \rangle$, as a function of the expected duration $t_2$ for two different values of $t_1$. Here, we focus on a scenario where option $A$ has lower intrinsic value but higher social influence strength. The results show that when the expected duration in the network $G_1$ is fixed at $t_1=10^2$, increasing the expected duration $t_2$ in the network $G_2$ leads to a monotonic increase in $\langle f_A \rangle$. When $t_2$ is small, the system spends most of the time in the sparse network $G_1$. As demonstrated in Fig.~\ref{t3}, a sparse network with a low average degree fails to amplify the difference in social appeal between options $A$ and $B$, resulting in a lower value of $\langle f_A \rangle$. As $t_2$ increases, the system spends more time in the network $G_2$, which enhances the disparity in social appeal between the two options, thereby further promoting $\langle f_A \rangle$. Additionally, when $t_{1}$ is large, achieving the same target value of $\langle f_A \rangle$ requires a proportionally larger $t_{2}$.

\section{Discussion}\label{Xsec11-5}\label{section5}
In this study, we construct an evolutionary model of collective decision-making to investigate the role of social influence in collective outcomes in static and dynamic networks. In our model, an individual's evaluation of an option is determined by two components: the option's intrinsic value and the social influence from neighbors. We propose a perceived utility function integrating these two aspects. Specifically, we assume that the social influence produced by neighbors adopting one option depends on the option's social influence strength and the number of neighbors who have adopted the same option.  By theoretically deriving the average frequency of option $A$, we find that social influence can amplify the advantage of the superior option or compensate for the deficiency of the inferior one. Specifically, when option $A$ holds a lower intrinsic value than that of option $B$ and option $B$ exerts greater social influence strength, the dominance of $B$ becomes an inevitable outcome. However, if option $A$ has stronger social influence strength, this effect can counteract its intrinsic value disadvantage, making individuals perceive $A$ as more popular and shifting the collective preference toward $A$.

Furthermore, we find that in static unweighted networks, the average degree exerts a dual effect on collective decision outcomes. Even if option $A$ is intrinsically inferior to $B$, its average frequency will increase with the network's average degree, provided that its social influence strength is stronger. Conversely, if the social appeal of option $A$ is weaker, a higher average degree further amplifies this disadvantage, reducing $A$'s frequency. This result is validated through computer simulations across various network structures (regular, small-world, scale-free, and random networks). Moreover, we consider our evolutionary model in dynamic networks, exploring how collective decision-making outcomes are affected when the population structure itself changes. We observe results similar to those in static networks. In addition, we find that when the system switches between different network topologies, the collective decision-making outcome depends on both the average degree of each network and the expected duration spent in each network configuration. The longer the system resides in a network with a high average degree, the more pronounced the social influence disparity becomes, as the higher average degree enhances its effect. More broadly, our model quantifies the fusion process of heterogeneous information sources-objective intrinsic value and local social information. The findings reveal that collective outcomes depend on network connectivity and the duration of each network configuration. This provides actionable regulatory tools for distributed cyber-physical systems and theoretical support for designing efficient collective decision-making protocols~\cite{bib0013}.

 Indeed, in our model the perceived utility of an individual $i$ can be viewed as the payoff obtained by playing a two-player, two-strategy game with neighbors on graphs. The payoff matrix for the game is accordingly given by
\begin{align*}
\left(\begin{matrix}\frac{\Pi_A}{d_i}+k_A\Pi_A&\frac{\Pi_A}{d_i}\\\frac{\Pi_B}{d_i}&\frac{\Pi_B}{d_i}+k_B\Pi_B\end{matrix}\right).
\end{align*}
Thus, when $\frac{\Pi_A-\Pi_B}{d_i}+k_A\Pi_A>0$ and $\frac{\Pi_B-\Pi_A}{d_i}+k_B\Pi_B>0$, our model can be transformed into a coordination game on graphs. In such a context, all players play the coordination game with the same payoff matrix on homogeneous graphs, but they engage in the coordination game with different payoff matrices on heterogeneous graphs since $d_i$ varies with nodes there. If the above inequality conditions are not satisfied, our model is transformed into other forms of games, such as, prisoner's dilemma game, depending on the parameter values. Thus such a transformation can help readers understand our work from the perspective of evolutionary games on graphs.

Recently, understanding how collective behavior emerges in a large population of individuals has become a popular research field~\cite{bib0037,bib0038,bib0039,bib0040,bib0041,bib0042,bib0043,bib0044,bib0045,bib0046,bib0047,bib0048}. In this context, relevant investigations have focused on how microscopic updating procedures influence the evolutionary outcomes at the population level. Our present study is closely related to such research. Indeed, a number of update rules
have been developed to mimic how agents make decisions, such as birth-death~\cite{bib0026,bib0037}, death-birth~\cite{bib0043,bib0044,bib0045}, and imitation~\cite{bib0038,bib0041,bib0046,bib0049,bib0050,bib0051,bib0052}. Differently, our model employs a different introspective dynamic mechanism, wherein individuals make decisions based on self-assessments of the perceived utility of the available options. This introspective mechanism reflects a cognitive process involving the evaluation of social information before a decision is made. Such an approach hence differentiates our model from other learning frameworks~\cite{bib0017,bib0018,bib0019,bib0053}, where individual learning (e.g., imitation based on intrinsic value) and social learning (e.g., conformist behavior) are treated as two independent processes. Under the introspective mechanism, we integrate an option's intrinsic value and social influence from neighbors into a unified perceived utility, providing a framework for understanding the interplay between network structures and social influence on collective decision-making. This framework may mirror real-world decision-making scenarios more vividly, where individuals are often indirectly influenced by others, leading to the formation of subjective judgments. In such scenarios, the effect of social influence may depend not only on how many neighbors have adopted one option, but also on the intrinsic value of the option itself. In particular, in the absence of social influence ($k_A = k_B = 0$), we reveal that the option with higher intrinsic value prevails in the population, which is also found in previous work~\cite{bib0017}. Conversely, in the presence of social influence, the option with lower intrinsic value may prevail, provided that it exerts a higher social influence strength.

Although introspection dynamics provide a rational mechanism for capturing individuals' self-assessment of perceived utility, the rule represents one particular microscopic updating procedure. In real-world scenarios, individuals may adopt diverse learning rules. Some individuals may evaluate options on their own~\cite{bib0025,bib0035}, whereas others may update their choices through imitation~\cite{bib0019}. Therefore, future work could extend the present framework by incorporating multiple learning mechanisms into a unified model. Such extensions would help investigate how social influence interacts with different learning mechanisms to shape collective decision-making outcomes. In addition, we focus on collective decision-making with binary options because it provides a baseline setting in which the interplay between intrinsic option values, local social influence, and network structure can be analyzed explicitly. However, many realistic decision-making scenarios involve more than two options. Extending the model to the framework of multi-option decision-making would allow us to predict collective decision-making outcomes in the realistic world. In addition, recent studies have examined how communication, emotional mechanisms, reputation-based evaluations, and social preferences influence cooperative behavior~\cite{bib0049,bib0050,bib0051,bib0052}. These factors indeed provide useful insight for characterizing social influence in collective decision-making. Future research could further extend the present framework by considering these factors to characterize social influence from various perspectives, thereby examining how such types of social influence shape collective decision-making outcomes in structured populations.

Moreover, the changes in the dynamic network in our model are assumed to be exogenous, meaning that network transition is independent of  individual decision-making process. In real social systems, however, the changes in network structures may depend on the option state and often coevolve with decision-making outcomes~\cite{bib0054,bib0055}. This coevolution of structure and behavior would, in turn, affect the impact of social influence~\cite{bib0054}. Additionally, our model describes local social influence through a mode of pairwise interaction. This provides a foundational and analytically feasible framework for depicting how the effects of neighbors' choices are integrated into an individual's perceived utility. However, in realistic scenarios, group interactions and higher-order interactions are often present~\cite{bib0056,bib0057,bib0058,bib0059}, and such interaction patterns can change over time. Therefore, future research could incorporate social influence involving group interactions and higher-order temporal interactions associated with complex contagion into the present framework, thereby exploring the evolutionary dynamics of collective decision-making with more realistic social interaction patterns.

\appendix
\setcounter{figure}{0}
\counterwithin*{figure}{part}
\renewcommand{\thefigure}{A.\arabic{figure}}
\setcounter{equation}{0}
\renewcommand{\theequation}{A.\arabic{equation}}
\section*{Appendix A}

\section*{Derivation details for Eq.~\eqref{x_expre}}

In the following, we prove that Eq.~\eqref{x_expre} is the unique solution to Eq.~\eqref{xM=x} by demonstrating that it satisfies $x_{\mathbf{s}} \mathbf{M}_{\mathbf{s}, \mathbf{s}^{\prime}} = x_{\mathbf{s}^{\prime}} \mathbf{M}_{\mathbf{s}^{\prime}, \mathbf{s}}$~\cite{bib0025}. The equation is obviously satisfied if $\Vert \mathbf{s}-\mathbf{s}^\prime \Vert_{1} \neq 1$. Specifically, if $\Vert \mathbf{s}-\mathbf{s}^\prime \Vert_{1} = 0$, we have $\mathbf{s}=\mathbf{s}^\prime$. The equation $x_{\mathbf{s}} \mathbf{M}_{\mathbf{s}, \mathbf{s}^{\prime}} = x_{\mathbf{s}^{\prime}} \mathbf{M}_{\mathbf{s}^{\prime}, \mathbf{s}}$ is evidently valid. If $\Vert \mathbf{s}-\mathbf{s}^\prime \Vert_{1} > 1$, both transition probabilities $\mathbf{M}_{\mathbf{s}, \mathbf{s}^{\prime}}$ and $\mathbf{M}_{\mathbf{s}^{\prime}, \mathbf{s}}$ are zero. It indicates that $x_{\mathbf{s}} \mathbf{M}_{\mathbf{s}, \mathbf{s}^{\prime}} = x_{\mathbf{s}^{\prime}} \mathbf{M}_{\mathbf{s}^{\prime}, \mathbf{s}}=0$. For adjacent states where $\Vert \mathbf{s}-\mathbf{s}^\prime \Vert_{1} = 1$, let $\mathbf{s} = (s_i, \mathbf{s}_{-i})$ and $\mathbf{s}^{\prime} = (s_i^{\prime}, \mathbf{s}_{-i})$ with $i=l(\mathbf{s}, \mathbf{s}^{\prime})$.  The detailed balance condition can be written as
\begin{align}\label{radio}
\frac{x_{\mathbf{s}}}{x_{\mathbf{s}^{\prime}}} = \frac{\mathbf{M}_{\mathbf{s}^{\prime}, \mathbf{s}}}{\mathbf{M}_{\mathbf{s}, \mathbf{s}^{\prime}}}.
\end{align}
Calculating the right-hand side of Eq.~\eqref{radio}, we have
\begin{align*}
\frac{\mathbf{M}_{\mathbf{s}^{\prime}, \mathbf{s}}}{\mathbf{M}_{\mathbf{s}, \mathbf{s}^{\prime}}} &= \frac{\hat{P}(\mathbf{s}^{\prime},\mathbf{s})}{\hat{P}(\mathbf{s},\mathbf{s}^\prime)} = \frac{\frac{1}{N} \left[s_i^{\prime}P_{A\rightarrow B}(\mathbf{s}_{-i})+(1-s_i^{\prime})P_{B\rightarrow A}(\mathbf{s}_{-i})\right]}{\frac{1}{N} \left[s_{i}P_{A\rightarrow B}(\mathbf{s}_{-i})+(1-s_{i})P_{B\rightarrow A}(\mathbf{s}_{-i})\right]}\\
&=\frac{s_i^{\prime}+(1-s_i^{\prime}) e^{\beta(U_{i}\left(A, \mathbf{s}_{-i}\right)-U_{i}\left(B, \mathbf{s}_{-i}\right))}}{s_i+(1-s_i)e^{\beta(U_{i}\left(A, \mathbf{s}_{-i}\right)-U_{i}\left(B, \mathbf{s}_{-i}\right))}}\\
&=\frac{ e^{\beta(1-s_i^{\prime})(U_{i}\left(A, \mathbf{s}_{-i}\right)-U_{i}\left(B, \mathbf{s}_{-i}\right))}}{e^{\beta(1-s_i)(U_{i}\left(A, \mathbf{s}_{-i}\right)-U_{i}\left(B, \mathbf{s}_{-i}\right))}}\\
&= e^{\beta\left(\hat{U}_{i}\left(s_{i}, \mathbf{s}_{-i}\right)-\hat{U}_{i}\left(s_{i}^{\prime}, \mathbf{s}_{-i}\right)\right)},
\end{align*}
where $\hat{U}_{i}(s_i,\mathbf{s}_{-i})=
s_i U_i(A,\mathbf{s}_{-i})+(1-s_i)U_i(B,\mathbf{s}_{-i})$.
Similarly, we have
\begin{align*}
\frac{x_{\mathbf{s}}}{x_{\mathbf{s}^{\prime}}} = \frac{e^{\beta \phi(\mathbf{s})}}{e^{\beta \phi(\mathbf{s}^{\prime})}} = e^{\beta (\phi(s_i, \mathbf{s}_{-i}) - \phi(s_{i}^{\prime}, \mathbf{s}_{-i}))}.
\end{align*}
Thus, Eq.~\eqref{radio} holds if
\begin{align}\label{U_phi}
\hat{U}_{i}(s_{i}, \mathbf{s}_{-i})-\hat{U}_{i}(s_{i}^{\prime}, \mathbf{s}_{-i})=\phi(s_i, \mathbf{s}_{-i}) - \phi(s_{i}^{\prime}, \mathbf{s}_{-i}).
\end{align}
Next, we prove that the given function $\phi$ in our model satisfies Eq.~\eqref{U_phi}. Note that
\begin{align*}
\phi(\mathbf{s}) =~& N \Pi_{B}+\sum_{i}s_i\left(\Pi_{A}-\Pi_{B}\right)+\frac{1}{2}k_{A} \Pi_{A} \sum_{i}\sum_{j} s_{i} s_{j} \omega_{ij} \\
&+\frac{1}{2}k_{B} \Pi_{B} \sum_{i}\sum_{j}\left(1-s_{i}\right)\left(1-s_{j}\right) \omega_{ij}.
\end{align*}
We then decompose $\phi(\mathbf{s})$ into terms dependent on $s_k$ and terms independent of $s_k$. Specifically, for a state $\mathbf{s}$, the function $\phi(\mathbf{s})$ can be decomposed into two distinct components: $\phi_1(\mathbf{s}_{-k})$ and $ \phi_2(s_k, \mathbf{s}_{-k})$. We have $\phi(\mathbf{s}) = \phi_1(\mathbf{s}_{-k}) + \phi_2(s_k, \mathbf{s}_{-k})$. Here, $\phi_1(\mathbf{s}_{-k})$ includes all terms that are strictly independent of individual $k$'s option state $s_k$, and is given by
\begin{align*}
\phi_1(\mathbf{s}_{-k}) =~& N \Pi_{B}+\sum_{i \neq k} s_{i}\left(\Pi_{A}-\Pi_{B}\right)+\frac{1}{2}k_{A} \Pi_{A} \sum_{i \neq k}\sum_{j\neq k} s_{i}s_{j} \omega_{ij} \\
& +\frac{1}{2}k_{B} \Pi_{B} \sum_{i \neq k}\sum_{j\neq k}\left(1-s_{i}\right)\left(1-s_{j}\right) \omega_{ij}.
\end{align*}
Besides, $\phi_2(s_k, \mathbf{s}_{-k})$ contains all terms involving $s_k$, given by
\begin{align*}
\phi_2(s_k, \mathbf{s}_{-k}) =~& s_{k}\left(\Pi_{A}-\Pi_{B}\right) +\frac{1}{2}k_{A} \Pi_{A}\Biggl(s_{k} \sum_{j}s_{j} \omega_{kj}+\sum_{i}s_{i}s_{k} \omega_{ik}\Biggr) \\
& +\frac{1}{2}k_{B} \Pi_{B}\left[\left(1-s_{k}\right) \sum_{j}\left(1-s_{j}\right) \omega_{k j}+\sum_{i}\left(1-s_{i}\right)\left(1-s_{k}\right)\omega_{ik}\right].
\end{align*}
Since the network $G$ is undirected, the adjacency matrix is symmetric ($\omega_{ij}=\omega_{ji}$). We can combine the symmetric sums (e.g., $\sum_{j} s_j \omega_{kj} = \sum_{i} s_i \omega_{ik}$) and simplify $\phi_2$ as follows
\begin{align*}
\phi_{2}\left(s_{k}, \mathbf{s}_{-k}\right) =~& s_{k}\left(\Pi_{A}-\Pi_{B}\right)+\frac{1}{2} k_{A} \Pi_{A}\left(2 s_{k} \sum_{j\neq k}s_{j} \omega_{k j}\right) \\
& +\frac{1}{2} k_{B} \Pi_{B}\left[2\left(1-s_{k}\right) \sum_{j\neq k}\left(1-s_{j}\right) \omega_{kj}\right] \\
=~& s_{k}\left(\Pi_{A}-\Pi_{B}\right) +k_{A} \Pi_{A} s_{k} n_{A}\left(\mathbf{s}_{-k}\right)+k_{B} \Pi_{B}\left(1-s_{k}\right) n_{B}\left(\mathbf{s}_{-k}\right).
\end{align*}

Then, we consider that individual $k$ switches $s_k$ to $s_k^{\prime}$. The resulting change in the function $\phi$ depends only on $\phi_2$, as $\phi_1$ remains constant. Accordingly, we have
\begin{align*}
\phi \left(s_{k}, \mathbf{s}_{-k}\right)-\phi\left(s_{k}^{\prime}, \mathbf{s}_{-k}\right)=~& \phi_{2}\left(s_{k}, \mathbf{s}_{-k}\right)-\phi_{2}\left(s_{k}^{\prime}, \mathbf{s}_{-k}\right) \\
=~& \left(s_{k}-s_{k}^{\prime}\right)\left(\Pi_{A}-\Pi_{B}\right) +\left(s_{k}-s_{k}^{\prime} \right) k_{A} \Pi_{A} n_{A}\left(\mathbf{s}_{-k}\right) \\
& +\left[\left(1-s_{k}\right) - \left(1-s_{k}^{\prime}\right)\right] k_{B} \Pi_{B} n_{B}\left(\mathbf{s}_{-k}\right)\\
=~& s_{k}\left[\Pi_{A}+k_{A} \Pi_{A} n_{A}\left(\mathbf{s}_{-k}\right)\right] - s_{k}^{\prime} \left[\Pi_{A}+k_{A} \Pi_{A} n_{A}\left(\mathbf{s}_{-k}\right)\right] \\
& + (1-s_{k})\left[\Pi_{B}+k_{B} \Pi_{B} n_{B}\left(\mathbf{s}_{-k}\right)\right]\\
&- (1-s_{k}^{\prime})\left[\Pi_{B}+k_{B} \Pi_{B} n_{B}\left(\mathbf{s}_{-k}\right)\right].
\end{align*}
Then, substituting Eq.~\eqref{U} into the above equation, we have
\begin{align*}
\phi \left(s_{k}, \mathbf{s}_{-k}\right)-\phi\left(s_{k}^{\prime}, \mathbf{s}_{-k}\right)=~& \left[ s_{k} U_{k}\left(A, \mathbf{s}_{-k}\right) + (1-s_{k}) U_{k}\left(B, \mathbf{s}_{-k}\right) \right] \\
& - \left[ s_{k}^{\prime} U_{k}\left(A, \mathbf{s}_{-k}\right) + (1-s_{k}^{\prime}) U_{k}\left(B, \mathbf{s}_{-k}\right) \right] \\
=~& \hat{U}_{k}\left(s_{k}, \mathbf{s}_{-k}\right)-\hat{U}_{k}\left(s_{k}^{\prime}, \mathbf{s}_{-k}\right).
\end{align*}
Thus, Eq.~\eqref{U_phi} holds, thereby validating Eq.~\eqref{radio}. In summary, we have $x_{\mathbf{s}} \mathbf{M}_{\mathbf{s}, \mathbf{s}^{\prime}} = x_{\mathbf{s}^{\prime}} \mathbf{M}_{\mathbf{s}^{\prime}, \mathbf{s}}$.

Furthermore, let $(\mathbf{x M})_{\mathbf{s}^{\prime}}$ denote the total probability of transitioning from all possible previous states $\mathbf{s}$ to the current state $\mathbf{s}^{\prime}$. Using the equation $x_{\mathbf{s}} \mathbf{M}_{\mathbf{s}, \mathbf{s}^{\prime}} = x_{\mathbf{s}^{\prime}} \mathbf{M}_{\mathbf{s}^{\prime}, \mathbf{s}}$, we have
\begin{align*}
(\mathbf{x M})_{\mathbf{s}^{\prime}} &=\sum_{\mathbf{s} \in \mathcal{S}}{x_{\mathbf{s}} \mathbf{M}_{\mathbf{s}, \mathbf{s}^{\prime}}} \notag =\sum_{\mathbf{s} \in \mathcal{S}} x_{\mathbf{s}^{\prime}} \mathbf{M}_{\mathbf{s}^{\prime}, \mathbf{s}}= x_{\mathbf{s}^{\prime}} \sum_{\mathbf{s} \in \mathcal{S}} \mathbf{M}_{\mathbf{s}^{\prime}, \mathbf{s}} \notag \\
&=x_{\mathbf{s}^{\prime}}.
\end{align*}
Since $\mathbf{s}^{\prime}$ is arbitrary, the above equation implies $\mathbf{x M} = \mathbf{x}$. Furthermore, we have $x_{\mathbf{s}}>0$ and
\begin{align*}
\sum_{\mathbf{s} \in \mathcal{S}} x_{\mathbf{s}} &= \sum_{\mathbf{s} \in \mathcal{S}} \frac{e^{\beta \phi(\mathbf{s})}}{\sum_{\mathbf{s}^{\prime} \in \mathcal{S}} e^{\beta \phi\left(\mathbf{s}^{\prime}\right)}}= \frac{1}{\sum_{\mathbf{s}^{\prime} \in \mathcal{S}} e^{\beta \phi\left(\mathbf{s}^{\prime}\right)}} \sum_{\mathbf{s} \in \mathcal{S}} e^{\beta \phi(\mathbf{s})}\\
&= 1.
\end{align*}
Thus, Eq.~\eqref{x_expre} is a solution to Eq.~\eqref{xM=x}, i.e., it represents a stationary distribution of the Markov chain. Given that the Markov chain is finite, irreducible, and aperiodic, this stationary distribution is unique.

\section*{Proof of Theorem~\ref{Theorem 1}}

Given a selection intensity $\beta > 0$, the stationary distribution of state $\mathbf{s}$ is
\begin{align*}
x_{\mathbf{s}}=\frac{e^{\beta \phi(\mathbf{s})}}{\sum_{\mathbf{s}^{\prime}} e^{\beta \phi(\mathbf{s}^{\prime})}}.
\end{align*}

So the average frequency of option $A$ is given by
\begin{align*}
\bar{f_A}=\sum_{\mathbf{s}} f_{A}(\mathbf{s}) x_{\mathbf{s}}=\frac{\sum_{\mathbf{s}} f_{A}(\mathbf{s})e^{\beta \phi(\mathbf{s})}}{\sum_{\mathbf{s}^{\prime}} e^{\beta \phi(\mathbf{s}^{\prime})}}.
\end{align*}

Let
\begin{align*}
F_1(\beta)=\sum_{\mathbf{s}} f_{A}(\mathbf{s}) e^{\beta \phi(\mathbf{s})}~\text{and} \quad F_2(\beta)=\sum_{\mathbf{s}} e^{\beta \phi(\mathbf{s})}.
\end{align*}
Then, $\bar{f_A}$ can be written as the ratio $\bar{f_A}(\beta) = F_1(\beta) / F_2(\beta)$. Performing a Taylor expansion around $\beta = 0$, we obtain
\begin{align}\label{taylor_expansion}
\bar{f_A} = \frac{F_1(0)}{F_2(0)} + \beta \left( \frac{F_1'(0) F_2(0) - F_1(0) F_2'(0)}{\left(F_2(0)\right)^2} \right) + O(\beta^2).
\end{align}
Here
\begin{align*}
F_1(0)=&\sum_{\mathbf{s}} f_{A}(\mathbf{s})=\sum_{\mathbf{s}}\frac{1}{N} \sum_{i=1}^{N} s_{i}=\sum_{m=0}^{N}\binom{N}{m}\frac{m}{N}\\
=~&2^{N-1}
\end{align*}
and
\begin{align*}
F_1^{\prime}(0)=&\sum_{\mathbf{s}} f_{A}(\mathbf{s})\phi(\mathbf{s}),F_2(0)=\sum_{\mathbf{s}}1=2^{N},F_2^{\prime}(0)=\sum_{\mathbf{s}}\phi(\mathbf{s}).
\end{align*}
Thus, we have
\begin{align}\label{ya}
\bar{f_A}=\frac{1}{2}+\beta \frac{\sum_{\mathbf{s}} f_{A}(\mathbf{s})\phi(\mathbf{s})-\frac{1}{2} \sum_{\mathbf{s}}\phi(\mathbf{s})}{2^{N}}+O\left(\beta^{2}\right),
\end{align}
where
\begin{align}\label{linear_coeff}
\frac{d \bar{f_A}}{d \beta}\bigg|_{\beta=0}= \frac{1}{2^N} \left( \sum_{\mathbf{s}} f_{A}(\mathbf{s}) \phi(\mathbf{s}) - \frac{1}{2} \sum_{\mathbf{s}} \phi(\mathbf{s}) \right).
\end{align}

To simplify Eq.~\eqref{linear_coeff}, we can rewrite the term $\sum_{\mathbf{s}} f_{A}(\mathbf{s}) \phi(\mathbf{s}) - \frac{1}{2} \sum_{\mathbf{s}} \phi(\mathbf{s})$ as
\begin{align}
\sum_{\mathbf{s}} f_{A}(\mathbf{s}) \phi(\mathbf{s}) - \frac{1}{2} \sum_{\mathbf{s}} \phi(\mathbf{s})
&= \sum_{\mathbf{s}} \left( f_{A}(\mathbf{s}) - \frac{1}{2} \right) \phi(\mathbf{s}) \notag \\
&= \sum_{\mathbf{s}} \left[ \left( \frac{1}{N} \sum_{i} s_{i} \right) - \frac{1}{2} \right] \phi(\mathbf{s}) \notag \\
&= \frac{1}{N} \sum_{i} \left[ \sum_{\mathbf{s}} \left( s_{i} - \frac{1}{2} \right) \phi(\mathbf{s}) \right]. \label{sum_swap}
\end{align}

Focusing on the inner summation over $\mathbf{s}$ for a fixed individual $i$, we can partition the state space $\mathcal{S}$ into two subsets: $\mathcal{S}_{i=1} = \{\mathbf{s} \mid s_i=1\}$ and $\mathcal{S}_{i=0} = \{\mathbf{s} \mid s_i=0\}$. The term $(s_i - 1/2)$ takes the value $1/2$ when $s_i=1$ and $-1/2$ when $s_i=0$. Thus, the summation is decomposed as follows
\begin{align}
\sum_{\mathbf{s}} \left( s_{i} - \frac{1}{2} \right) \phi(\mathbf{s})
&= \sum_{\mathbf{s} \in \mathcal{S}_{i=1}} \left(1 - \frac{1}{2}\right) \phi(\mathbf{s})+ \sum_{\mathbf{s} \in \mathcal{S}_{i=0}} \left(0 - \frac{1}{2}\right) \phi(\mathbf{s}) \notag \\
&= \frac{1}{2} \left( \sum_{\mathbf{s} \in \mathcal{S}_{i=1}} \phi(\mathbf{s}) - \sum_{\mathbf{s} \in \mathcal{S}_{i=0}} \phi(\mathbf{s}) \right). \label{split_sums}
\end{align}

When the option state of individual $i$ is fixed, the summation over $\mathbf{s}$ can further be written as a summation over $\mathbf{s}_{-i}$. Then, Eq.~\eqref{split_sums} can be expressed as
\begin{align*}
\frac{1}{2}\left( \sum_{\mathbf{s} \in \mathcal{S}_{i=1}} \phi(\mathbf{s}) - \sum_{\mathbf{s} \in \mathcal{S}_{i=0}} \phi(\mathbf{s}) \right)=\frac{1}{2}\left(\sum_{\mathbf{s}_{-i}} \phi\left(1, \mathbf{s}_{-i}\right)-\sum_{\mathbf{s}_{-i}} \phi\left(0, \mathbf{s}_{-i}\right)\right).
\end{align*}

Let $\hat{\phi_i}=\frac{1}{2}\left(\sum_{\mathbf{s}_{-i}} \phi\left(1, \mathbf{s}_{-i}\right)-\sum_{\mathbf{s}_{-i}} \phi\left(0, \mathbf{s}_{-i}\right)\right)$. Furthermore, based on Eq.~\eqref{U_phi}, we obtain $\phi\left(1, \mathbf{s}_{-i}\right)-\phi\left(0, \mathbf{s}_{-i}\right) = U_{i}\left(A, \mathbf{s}_{-i}\right) - U_i(B, \mathbf{s}_{-i})$. Then, substituting the above equation into $\hat{\phi_i}$, we have
\begin{align}
\hat{\phi_i}&=\frac{1}{2} \sum_{\mathbf{s}_{-i}}\left(U_{i}\left(A, \mathbf{s}_{-i}\right)-U_{i}\left(B, \mathbf{s}_{-i}\right)\right)\notag\\
&=\frac{1}{2} \sum_{\mathbf{s}_{-i}}\left(\Pi_{A}-\Pi_{B} +n_{A}(\mathbf{s}_{-i})k_{A}\Pi_{A}-n_{B}(\mathbf{s}_{-i})k_{B}\Pi_{B}\right),\label{hatphi}
\end{align}
where
\begin{align*}
\sum_{\mathbf{s}_{-i}}\left(\Pi_{A}-\Pi_{B}\right)&=\sum_{m=0}^{N-1}\binom{N-1}{m}\left(\Pi_{A}-\Pi_{B}\right)\\
&=2^{N-1}\left(\Pi_{A}-\Pi_{B}\right),\\
\sum_{\mathbf{s}_{-i}} n_{A}\left(\mathbf{s}_{-i}\right)&=\sum_{\mathbf{s}_{-i}} \sum_{j\neq i} \omega_{ij} s_{j}\left(\mathbf{s}_{-i}\right)=\sum_{j\neq i} \omega_{ij} \sum_{\mathbf{s}_{-i}} s_{j}\left(\mathbf{s}_{-i}\right) \\
&=\sum_{j\neq i} \omega_{ij}\frac{1}{2}\sum_{m=0}^{N-1}\binom{N-1}{m}=\sum_{j\neq i} \omega_{ij} 2^{N-2}\\
&=d_{i} 2^{N-2},\\
\sum_{\mathbf{s}_{-i}} n_{B}\left(\mathbf{s}_{-i}\right)&=\sum_{\mathbf{s}_{-i}} \sum_{j\neq i} \omega_{ij}\left(1- s_{j}\left(\mathbf{s}_{-i}\right)\right)=\sum_{j\neq i} \omega_{ij} \sum_{\mathbf{s}_{-i}} \left(1- s_{j}\left(\mathbf{s}_{-i}\right)\right) \\
&=\sum_{j\neq i} \omega_{ij}\frac{1}{2}\sum_{m=0}^{N-1}\binom{N-1}{m}=\sum_{j\neq i} \omega_{ij} 2^{N-2}\\
&=d_{i} 2^{N-2}.
\end{align*}

Therefore, substituting the above expressions into $\hat{\phi_i}$, we have
\begin{align}
\hat{\phi_i} & =\frac{1}{2}\left[2^{N-1}\left(\Pi_{A}-\Pi_{B}\right)+d_{i} 2^{N-2}\left(k_{A} \Pi_{A}-k_{B} \Pi_{B}\right)\right]\notag\\
&=2^{N-2}\left[\left(\Pi_{A}-\Pi_{B}\right)+\frac{d_{i}}{2}\left(k_{A} \Pi_{A}-k_{B} \Pi_{B}\right)\right].\label{hatphi1}
\end{align}

Then, using Eqs.~\eqref{sum_swap} and~\eqref{hatphi1}, we obtain
\begin{align}
&\sum_{\mathbf{s}} f_{A}(\mathbf{s}) \phi(\mathbf{s})-\frac{1}{2} \sum_{\mathbf{s}} \phi(\mathbf{s})\notag\\
=~&\frac{1}{N}\sum_{i}\left\{2^{N-2}\left[\left(\Pi_{A}-\Pi_{B}\right)+\frac{d_{i}}{2}\left(k_{A} \Pi_{A}-k_{B} \Pi_{B}\right)\right]\right\}\notag\\
=~&2^{N-2}\left[\left(\Pi_{A}-\Pi_{B}\right)+\frac{\bar{d}}{2}\left(k_{A} \Pi_{A}-k_{B} \Pi_{B}\right)\right].\label{fn}
\end{align}

Finally, substituting Eq.~\eqref{fn} into Eq.~\eqref{ya}, we have
\begin{align*}
\bar{f_A}=~&\frac{1}{2}+\frac{\beta}{4} \left[\left(\Pi_{A} - \Pi_{B}\right) + \frac{\bar{d}}{2}\left(k_{A}\Pi_{A} - k_{B}\Pi_{B}\right)\right]+O\left(\beta^{2}\right).
\end{align*}

This completes the proof of Theorem~\ref{Theorem 1}.

\section*{Proof of Theorem~\ref{Theorem 2}}

We know that the average frequency of option $ A $ in dynamic networks is given by
\begin{align*}
\langle f_A \rangle=\sum_{\mathbf{s},\delta} z_{(\mathbf{s},\delta)}f_{A}\left(\mathbf{s}\right).
\end{align*}

According to Theorem~\ref{Theorem 1}, for any fixed network configuration $G_{\delta}$, the average frequency of option $A$ under the weak selection limit ($\beta \to 0$) is given by
\begin{align*}
\bar{f_A}^{(\delta)}=~&\frac12+\frac{\beta}{4}\left[(\Pi_{A}-\Pi_{B})+\frac{\bar{d}^{(\delta)}}{2}\,(k_A\Pi_{A}-k_B\Pi_{B})\right]+O(\beta^2),
\end{align*}
where $\bar{d}^{(\delta)} = \frac{1}{N} \sum_{i=1}^{N} d_{i}^{(\delta)}$ denotes the average weighted degree of network $G_{\delta}$. We subsequently demonstrate that $\langle f_A \rangle$ exhibits a functional form analogous to $\bar{f_A}^{(\delta)}$.

For any $ \mathbf{s}, \mathbf{s}^{\prime} \in \mathcal{S} $ and $ \gamma, \delta \in\mathcal{L} $, the joint transition probability satisfies
\begin{equation*}
\mathbf{H}_{(\mathbf{s},\gamma),(\mathbf{s}^{\prime},\delta)}=\mathbf{M}^{(\gamma)}_{\mathbf s,\mathbf{s}^{\prime}}q_{\gamma\delta}.
\end{equation*}
Since
\begin{align*}
\sum_{\mathbf{s}, \delta} z_{(\mathbf{s}, \delta)} f_{A}(\mathbf{s}) & =\sum_{\mathbf{s}^{\prime}, \gamma} z_{(\mathbf{s}^{\prime}, \gamma)}  f_{A}\left(\mathbf{s}^{\prime}\right) \\
& =\sum_{\mathbf{s}^{\prime}, \gamma} \sum_{\mathbf{s}, \delta} z_{(\mathbf{s}, \delta)} \mathbf{H}_{(\mathbf{s}, \delta),(\mathbf{s}^{\prime}, \gamma)}f_{A}\left(\mathbf{s}^{\prime}\right)\\ & =\sum_{\mathbf{s}^{\prime}, \gamma} \sum_{\mathbf{s}, \delta} z_{(\mathbf{s}, \delta)} \mathbf{M}^{(\delta)} _{\mathbf{s}, \mathbf{s}^{\prime}}q_{\delta \gamma} f_{A}\left(\mathbf{s}^{\prime}\right) \\
& =\sum_{\mathbf{s}^{\prime}} \sum_{\mathbf{s}, \delta} z_{(\mathbf{s}, \delta)} \mathbf{M}^{(\delta)}_{\mathbf{s}, \mathbf{s}^{\prime}} f_{A}\left(\mathbf{s}^{\prime}\right),
\end{align*}
we have
\begin{align}\label{zM_f}
\sum_{\mathbf{s}, \delta} z_{(\mathbf{s}, \delta)} \left(\sum_{\mathbf{s}^{\prime}} \mathbf{M}^{(\delta)}_{\mathbf{s}, \mathbf{s}^{\prime}} f_{A}\left(\mathbf{s}^{\prime}\right)-f_{A}(\mathbf{s})\right)=0.
\end{align}

Then, we perform a Taylor expansion of the transition probability $\mathbf{M}^{(\delta)}_{\mathbf{s}, \mathbf{s}^{\prime}}$ around $\beta=0$. Let $\Delta U_{j}^{(\delta)}\left(\mathbf{s}_{-j}\right)=U_{j}^{(\delta)}\left(A, \mathbf{s}_{-j}\right)-U_{j}^{(\delta)}\left(B, \mathbf{s}_{-j}\right)$. When $\beta\rightarrow 0$, we have
\begin{align*}
P_{B \rightarrow A}^{(\delta)}\left(\mathbf{s}_{-j}\right)=\frac{1}{1+e^{-\beta \Delta U_{j}^{(\delta)}}}=\frac{1}{2}+\frac{\beta}{4} \Delta U_{j}^{(\delta)}+O\left(\beta^{2}\right), \\
P_{A \rightarrow B}^{(\delta)}\left(\mathbf{s}_{-j}\right)=\frac{1}{1+e^{+\beta \Delta U_{j}^{(\delta)}}}=\frac{1}{2}-\frac{\beta}{4} \Delta U_{j}^{(\delta)}+O\left(\beta^{2}\right).
\end{align*}

Let $\mathcal{N}(\mathbf{s})=\left\{\mathbf{s}^{\prime} \mid\left\|\mathbf{s}-\mathbf{s}^{\prime}\right\|_{1}=1\right\}$.  Then, for any $\mathbf{s}^{\prime}\in\mathcal{N}(\mathbf{s})$, we have
\begin{align}
\mathbf{M}_{\mathbf{s}, \mathbf{s}^{\prime}}^{(\delta)}&=\frac{1}{N}\left[s_{i} P_{A \rightarrow B}^{(\delta)}\left(\mathbf{s}_{-i}\right)+\left(1-s_{i}\right) P_{B \rightarrow A}^{(\delta)}\left(\mathbf{s}_{-i}\right)\right]\notag\\
&=\frac{1}{2N}+\frac{\beta}{4N}\left(1-2 s_{i}\right) \Delta U_{i}^{(\delta)}\left(\mathbf{s}_{-i}\right)+O\left(\beta^{2}\right)\label{M_P}
\end{align}
and
\begin{align}f_A(\mathbf{s}^{\prime})-f_A(\mathbf{s})=\frac{1}{N}(1-2s_i).\label{fi_f}
\end{align}
Here $i=l(\mathbf{s}, \mathbf{s}^{\prime})$. Thus, according to Eqs.~(\ref{M_P}) and (\ref{fi_f}), we have
\begin{align}
\sum_{\mathbf{s}^{\prime}}\mathbf{M}_{\mathbf{s},\mathbf{s}^{\prime}}^{(\delta)}f_A(\mathbf{s}^{\prime})-f_A(\mathbf{s})&=\sum_{\mathbf{s}^{\prime}\in\mathcal{N}(\mathbf{s})}\mathbf{M}_{\mathbf{s},\mathbf{s}^{\prime}}^{(\delta)}\big[f_A(\mathbf{s}^{\prime})-f_A(\mathbf{s})\big]\notag\\
&=\frac{1}{N}\left(\frac{1}{2}-f_A(\mathbf{s})\right)+\frac{\beta}{4N^2}\sum_{i=1}^N\Delta U_i^{(\delta)}(\mathbf{s}_{-i})+O(\beta^2).\label{mf_f}
\end{align}
Thus, substituting Eq.~(\ref{mf_f}) into Eq.~\eqref{zM_f}, we obtain
\begin{align}\label{zM_f_2}
&\sum_{\mathbf{s}, \delta} z_{(\mathbf{s}, \delta)} \Biggl[\frac{1}{N}\left(\frac{1}{2}-f_A(\mathbf{s})\right)+\frac{\beta}{4N^2}\sum_{i=1}^N\Delta U_i^{(\delta)}(\mathbf{s}_{-i})+O(\beta^2)\Biggr]=0.
\end{align}

To further simplify Eq.~\eqref{zM_f_2}, we define the marginal distribution over network states as $ \mathbf{m}=\left(m_{\delta}\right)_{\delta \in \mathcal{L}} $, where $ m_{\delta}=\sum_{\mathbf{s}} z_{(\mathbf{s}, \delta)} $. Using the condition $ \mathbf{z}=\mathbf{z H} $, it follows that for any $ \delta \in \mathcal{L} $,
\begin{align*}
m_\delta=\sum_{\mathbf{s}^{\prime}}z_{(\mathbf{s}^{\prime},\delta)}
&=\sum_{\mathbf{s}^{\prime}, \mathbf {s}, \gamma}z_{(\mathbf s,\gamma)}\,\mathbf{M}^{(\gamma)}_{\mathbf s,\mathbf{s}^{\prime}}\,q_{\gamma\delta}\\
&=\sum_{\gamma}q_{\gamma\delta}\sum_{\mathbf s}z_{(\mathbf s,\gamma)}
\Big(\sum_{\mathbf{s}^{\prime}}\mathbf{M}^{(\gamma)}_{\mathbf{s},\mathbf{s}^{\prime}}\Big)\\
&=\sum_{\gamma}q_{\gamma\delta}\,m_\gamma,
\end{align*}
which implies that the vector $ \mathbf{m} $ satisfies
$$
\mathbf{m Q}=\mathbf{m}.
$$

Given the uniqueness of the stationary distribution of $ \mathbf{Q} $, we conclude that $ \mathbf{m}=\mathbf{y} $. Then, we have
\begin{align}\label{zzzy}
z_{(\mathbf{s},\delta)}=\sum_{\mathbf{s}^{\prime}}z_{(\mathbf{s}^{\prime},\delta)}\frac{z_{(\mathbf{s},\delta)}}{\sum_{\mathbf{s}^{\prime}}z_{(\mathbf{s}^{\prime},\delta)}} =y_{\delta} z_{\mathbf{s}|\delta},
\end{align}
where $\sum_{\mathbf{s}}z_{\mathbf{s}|\delta}=1$. Let $z_{\mathbf{s}|\delta}=z_{\mathbf{s}|\delta}(\beta)$. When $\beta=0$, the stationary distribution component corresponding to each state $\mathbf{s}$ is $2^{-N}$ in a given network $G_{\delta}$. Then, the Taylor expansion of $z_{\mathbf{s}|\delta}(\beta)$ at $\beta=0$ is given by
\begin{align}\label{z2b}
z_{\mathbf{s}|\delta}=2^{-N}+\beta\zeta_{\delta}(\mathbf{s})+O(\beta^{2}),
\end{align}
where $\zeta_{\delta}(\mathbf{s})=\left.\frac{\mathrm{d}z_{\mathbf{s}|\delta}}{\mathrm{d}\beta}\right|_{\beta =0}$ and $\sum_{\mathbf{s}}\zeta_{\delta}(\mathbf{s})=0$. Then, substituting Eqs.~(\ref{zzzy}) and (\ref{z2b}) into Eq.~(\ref{zM_f_2}), we have
\begin{align}
0=&\sum_{\delta}y_{\delta}\left[\frac{1}{N}\sum_{\mathbf{s}}\zeta_{\delta}(\mathbf{s})\left(\frac{1}{2}-f_{A}(\mathbf{s})\right)+\frac{1}{4N^2}\sum_{\mathbf{s}}2^{-N}\sum_{i}\Delta U_{i}^{(\delta)}(\mathbf{s}_{-i})\right]\beta+O(\beta^2).\label{0_y}
\end{align}
Since $\sum_{\mathbf{s}}\zeta_{\delta}(\mathbf{s})=0$, we have
\begin{align*}
\sum_\mathbf{s}\zeta_\delta(\mathbf{s})\left(\frac{1}{2}-f_A(\mathbf{s})\right)=-\sum_\mathbf{s}\zeta_\delta(\mathbf{s})f_A(\mathbf{s}).
\end{align*}
Thus, Eq.~\eqref{0_y} is
equivalent to
\begin{align*}
\beta\sum_\delta y_\delta\sum_\mathbf{s}\zeta_\delta(\mathbf{s})f_A(\mathbf{s})=~&\frac{\beta}{2^{N}4N}\sum_\delta y_\delta\sum_\mathbf{s}\sum_i\Delta U_i^{(\delta)}(\mathbf{s}_{-i})+O(\beta^2).
\end{align*}

As a result, when $\beta \rightarrow 0$, we obtain
\begin{align}
\langle f_A\rangle &= \sum_{\mathbf{s},\delta} f_A\left(\mathbf{s}\right) z_{(\mathbf{s},\delta)}\notag\\
&= \sum_{\delta} y_{\delta} \sum_{\mathbf{s}} f_{A}(\mathbf{s}) z_{\mathbf{s} \mid \delta}\notag\\
&= \sum_{\delta} y_{\delta}\sum_{\mathbf{s}} f_{A}(\mathbf{s}) (2^{-N}+\beta\zeta_{\delta}(\mathbf{s})+O(\beta^{2}))\notag\\
&= \sum_{\delta} y_{\delta}\sum_{\mathbf{s}} f_{A}(\mathbf{s})2^{-N} +\beta\sum_{\delta} y_{\delta}\sum_{\mathbf{s}} f_{A}(\mathbf{s})\zeta_{\delta}(\mathbf{s})+O(\beta^{2})\notag\\
&=2^{-N}\sum_{\delta} y_{\delta}\left(\sum_{\mathbf{s}} f_{A}(\mathbf{s}) +\frac{\beta}{4N}\sum_\mathbf{s}\sum_i\Delta U_i^{(\delta)}(\mathbf{s}_{-i})\right)+O(\beta^{2}).\label{dy_fa_2}
\end{align}

According to Eqs.~\eqref{hatphi} and~\eqref{hatphi1}, we have
\begin{align}
\frac{1}{N}\sum_\mathbf{s}\sum_i\Delta U_i^{(\delta)}(\mathbf{s}_{-i})&=\frac{2}{N}\sum_{i}\sum_{\mathbf{s}_{-i}}\Delta U_i^{(\delta)}(\mathbf{s}_{-i})\notag\\
&=2^{N}\frac{1}{N}\sum_i\Biggl[\left(\Pi_{A}-\Pi_{B}\right)+\frac{d_{i}^{(\delta)}}{2}\left(k_{A} \Pi_{A}-k_{B} \Pi_{B}\right)\Biggr]\notag\\
&=2^{N}\Biggl[\left(\Pi_{A}-\Pi_{B}\right)+\frac{\bar{d}^{(\delta)}}{2}\left(k_{A} \Pi_{A}-k_{B} \Pi_{B}\right)\Biggr].\label{NDU}
\end{align}

Finally, substituting Eq.~\eqref{NDU} into Eq.~\eqref{dy_fa_2}, we have
\begin{align*}
\langle f_A\rangle&= \sum_{\delta} y_{\delta}\Biggl\{\frac12+\frac{\beta}{4}\Bigl[
(\Pi_{A}-\Pi_{B})+\frac{\bar{d}^{(\delta)}}{2}\left(k_A\Pi_{A}
-\,k_B\Pi_{B}\right)\Bigr]+O(\beta^2)\Biggr\}\\
&=\frac12+\frac{\beta}{4}\Biggl[
\left(\Pi_{A}-\Pi_{B}\right)+\frac{1}{2}\left(\sum_{\delta} y_{\delta}\bar{d}^{(\delta)}\right)\left(k_A\Pi_{A}-\,k_B\Pi_{B}\right)\Biggr]+O(\beta^2)\\
&=\frac12+\frac{\beta}{4}\Biggl[
\left(\Pi_{A}-\Pi_{B}\right)+\frac{\langle d\rangle}{2}\left(k_A\Pi_{A}-k_B\Pi_{B}\right)\Biggr]+O(\beta^2),
\end{align*}
where $\langle d\rangle=\sum_{\delta=1}^{L} y_{\delta}\bar{d}^{(\delta)}=\sum_{\delta=1}^{L}\frac{t_\delta}{\sum_{k=1}^L t_k}\bar{d}^{(\delta)}$.

This completes the proof of Theorem~\ref{Theorem 2}.
\section*{CRediT authorship contribution statement}
\noindent \textbf{Yuyuan Liu:} Writing – review \& editing, Writing – original draft, Investigation, Formal analysis.
\\
\textbf{Xiaojie Chen:} Writing – review \& editing, Supervision, Formal analysis, Conceptualization.

\section*{Declaration of competing interest}
The authors declare that they have no known competing financial interests or personal relationships that could have appeared to influence the work reported in this paper.

\section*{Data availability}

No data was used for the research described in the article.

\section*{Acknowledgments}

This research was supported by the National Natural Science Foundation of China (Grant No. 62473081).







\end{document}